\documentclass{emulateapj}
\pdfoutput=1
\usepackage{apjfonts}
\usepackage{amssymb}
\usepackage{natbib}
 \usepackage{color}
\usepackage{graphicx,mathrsfs,amsmath}
\usepackage[caption=false]{subfig}    

\nonstopmode 

\usepackage{enumerate}
\newcommand{\kms}{$\rm{km\;s^{-1}}$}
\newcommand{\pcm}{$\rm{cm^{-2}}$}

\providecommand{\CIV}{\ensuremath{\mbox{\ion{C}{4}}}}
\providecommand{\CII}{\ensuremath{\mbox{\ion{C}{2}}}}
\providecommand{\SII}{\ensuremath{\mbox{\ion{S}{2}}}}
\providecommand{\SiII}{\ensuremath{\mbox{\ion{Si}{2}}}}
\providecommand{\SiIII}{\ensuremath{\mbox{\ion{Si}{3}}}}
\providecommand{\SiIV}{\ensuremath{\mbox{\ion{Si}{4}}}}
\providecommand{\AlII}{\ensuremath{\mbox{\ion{Al}{2}}}}
\providecommand{\NV}{\ensuremath{\mbox{\ion{N}{5}}}}
\providecommand{\HI}{\ensuremath{\mbox{\ion{H}{1}}}}
\providecommand{\OVI}{\ensuremath{\mbox{\ion{O}{6}}}}
\providecommand{\OI}{\ensuremath{\mbox{\ion{O}{1}}}}
\providecommand{\FeII}{\ensuremath{\mbox{\ion{Fe}{2}}}}
\providecommand{\NI}{\ensuremath{\mbox{\ion{N}{1}}}}

\shorttitle{Mapping the Nuclear Outflow of the Milky Way}
\shortauthors{Bordoloi et al.}

\begin{document}
\title{Mapping the Nuclear Outflow of the Milky Way: Studying the Kinematics and Spatial extent of the Northern Fermi Bubble}

\author{
Rongmon Bordoloi\altaffilmark{1,10}, 
Andrew J. Fox\altaffilmark{2}, 
Felix J. Lockman\altaffilmark{3}, 
Bart P. Wakker\altaffilmark{4},
Edward B. Jenkins\altaffilmark{5},
Blair D. Savage\altaffilmark{6},
Svea Hernandez\altaffilmark{7},
Jason Tumlinson\altaffilmark{2},
Joss Bland-Hawthorn\altaffilmark{8},
\&  Tae-Sun Kim\altaffilmark{9}
  }

\altaffiltext{1}{MIT-Kavli Center for Astrophysics and Space Research, 77 Massachusetts Avenue, Cambridge, MA, 02139, USA;\href{mailto:bordoloi@mit.edu}{bordoloi@mit.edu}}
\altaffiltext{2}{Space Telescope Science Institute, 3700 San Martin Drive, 21218, Baltimore, MD}
\altaffiltext{3}{Green Bank Observatory, P.O. Box 2, Rt. 28/92, Green Bank, WV 24944, USA}
\altaffiltext{4}{Supported by NASA/NSF; affiliated with Department of Astronomy, University of Wisconsin, Madison, 475 North Charter Street, Madison, WI 53706, USA}
\altaffiltext{5}{Princeton University Observatory, Princeton, NJ 08544, USA}
\altaffiltext{6}{Department of Astronomy, University of Wisconsin, Madison, 475 North Charter Street, Madison, WI 53706, USA}
\altaffiltext{7}{Department of Astrophysics/IMAPP, Radboud University Nijmegen, PO Box 9010, 6500 GL Nijmegen, The Netherlands}
\altaffiltext{8}{ Institute of Astronomy, School of Physics, University of Sydney, NSW 2006, Australia}
\altaffiltext{9}{ Osservatorio Astronomico di Trieste, Via G.B. Tiepolo 11, I-34143 Trieste, Italy}
\altaffiltext{10}{Hubble Fellow}


\begin{abstract}

We report new observations from a systematic, spectroscopic, ultraviolet absorption-line survey that maps the spatial and kinematic properties of the high-velocity gas in the Galactic Center region. We examine the hypothesis that this gas traces the biconical nuclear outflow. We use ultraviolet spectra of 47 background QSOs and halo stars projected inside and outside the northern Fermi Bubble from the \textit{Hubble Space Telescope} to study the incidence of high velocity absorption around it. We use five lines of sight inside the northern Fermi Bubble to constrain the velocity and column densities of outflowing gas traced by O I,  Al II, C II, C IV,  Si II, Si III, Si IV and other species. We find that all five lines of sight inside the northern Fermi Bubble exhibit blueshifted high velocity absorption components, whereas only 9 out of the 42 lines of sight outside the northern Fermi Bubble exhibit blueshifted high velocity absorption components. The observed outflow velocity profile decreases with Galactic latitude and radial distance ($R$) from the Galactic Center. The observed blueshifted velocities change from $v_{GSR} = -$265 {\kms} at  $R \approx$ 2.3 kpc to $v_{GSR} = -$91 {\kms} at  $R \approx$ 6.5 kpc. We derive the metallicity of the entrained gas along the 1H1613-097 sightline,  one that passes through the center of the northern Fermi Bubble, finding [O/H] $\gtrsim -0.54 \pm  0.15$. A simple kinematic model tuned to match the observed absorption component velocities along the five lines of sight inside the Bubble, constrains the outflow velocities to $\approx$ 1000 $-$ 1300 {\kms}, and the age of the outflow to be $\sim$ 6 $-$ 9 Myr. We estimate a minimum mass outflow rate for the nuclear outflow to be $\gtrsim$ 0.2 $\rm{ M_{\odot}\; yr^{-1}}$. Combining the age and mass outflow rates, we determine a minimum mass of total UV absorbing cool gas entrained in the Fermi Bubbles to be $\gtrsim \rm{ 2 \times 10^{6} M_{\odot}}$.  
\end{abstract}

\section{Introduction}
 In the modern picture of galaxy evolution, the exchange of gas between galaxies and their surrounding circumgalactic medium (CGM) plays a crucial role in establishing the properties of the galaxies. Feedback processes that regulate these exchanges of gas, are crucial in setting up the mass metallicity relation \citep{Tremonti2004}, the quenching of star-formation in massive galaxies \citep{Tremonti2007, Tripp2011}, and in explaining the mismatch between the galaxy stellar mass  function and the dark matter halo mass function \citep{Oppenheimer2010}. These powerful galactic outflows also must suppress the inflow of gas onto galaxies, constraining the assembly of the baryonic component and regulating the star-formation in galaxies \citep{dav11b,FaucherGiguere2011}.

The diffuse gas in the CGM can be detected as absorption lines in the continua of background quasar spectra (e.g. \citealt{Steidel1994,Bowen1995,Chen1998,Steidel2002,Wakker2009, Chen2010a,Stocke2013,Tumlinson2013,zhu2013a,Bordoloi2014c}). The outflowing gas can be detected as blueshifted absorption imprinted on the stellar continuum of the host galaxies themselves \citep{Weiner2009,Heckman2015,Bordoloi2016a}, giving the ``down-the-barrel'' view onto the galaxy in question, or of background galaxies offset by some impact parameter \citep{Steidel2010,Bordoloi2011a}. These galactic outflows are primarily biconical in morphology; both in nearby star-forming galaxies (see reviews by \citealt{Heckman2002} and  \citealt{Veilleux2005}), and in high redshift galaxies \citep{Bordoloi2014b,Rubin2014a}. However, all of these observational studies suffer from a major barrier that limits what we can learn from them about galaxy-gas flows: they use statistical sampling of one sightline for each of a sample of galaxies.

Our vantage point inside the disk of the Milky Way gives us a unique opportunity to break this deadlock and study the outflowing gas from the Milky Way itself, along multiple lines of sight.  Using multiple background sources to study the nuclear outflow of the Milky Way offers us a front row seat in understanding these feedback processes in unprecedented detail.

The Fermi Bubbles (FBs) \citep{Dobler2010,Su2010} are giant 1-100 GeV, $\gamma$ ray emitting structures that extend up to $\approx \pm$55$^{\circ}$ above and below the Galactic Center. These structures show enhanced emission in multiple wavelength ranges. Spatially coherent emission features have been observed in hard X-ray emission out to $l \approx 20^{\circ}$  \citep{,Bland-Hawthorn2003}, soft X-ray emission at their base (0.3-1.0 keV; \citealt{Snowden1997,Kataoka2013}), K-band microwave emission (23-94 GHz;  \citealt{Finkbeiner2004, Dobler2008}), and polarized radio emission at 2.3 GHz (synchrotron radiation; \citealt{Carretti2013}). \cite{Lockman1984} also reported a lack of 21 cm emission H I clouds near the center of the Milky Way and suggested that this might be cleared away by a wind. New 21 cm emission surveys towards the Galactic center have further provided evidence for such a wind \citep{2016arXiv160501140L}.

The energetic origin and the source of the $\gamma$-ray emission mechanism  that illuminate the Fermi Bubbles are still being debated today. There are two possible scenarios put forward to explain the origin of the Fermi Bubbles. One scenario argues that the Fermi Bubbles are the result of a recent explosive outburst from the central supermassive black hole of the Milky Way that happened a few Myrs ago, and the observed  $\gamma$ ray emission originates from inverse Compton scattering of a non thermal leptonic population \citep{Su2010,Zubovas2011,Fujita2013}. The second scenario argues that the Fermi bubbles originate from the integrated effect of secular processes taking place in the inner part of the Milky Way \citep{Thoudam2013} or Galactic nucleus, such as tidal disruption events regularly taking place every $10^4-10^5$ yr \citep{Cheng2011} or the continuous and vigorous star formation activity around the 200$-$300 pc diameter region around the central black hole of the Milky Way \citep{Lacki2014}, and the observed $\gamma$-ray emission is owing to hadronic collisions experienced by heaver ions and a population of cosmic-ray protons \citep{2011PhRvL.106j1102C}   (see \citealt{Crocker2015} for a discussion on the two scenarios). 

Discriminating between any of these processes would require knowledge of the kinematics of the Fermi Bubbles. Knowing the kinematics would allow us to independently constrain the age and spatial and kinematic extent of the cool entrained material inside the Fermi Bubbles. However to date, no study has systematically mapped out the kinematics and spatial extent of the possible nuclear outflow from the Milky Way. Only a handful of individual lines of sight have been used to trace the kinematics, ionization state and elemental abundance of the nuclear outflow \citep{Keeney2006,Bowen2008,Zech2008, Fang2014}. 

We have been conducting a survey with the \textit{Hubble Space Telescope} (HST) to systematically probe the kinematics and physical properties of the warm and cool diffuse gas in the Galactic Center (GC) region in absorption with UV spectroscopy (Program IDs (PID) 12936 and 13448). In a previous paper \citep{Fox2015}, we reported the discovery of high velocity gas components consistent with a biconical nuclear outflow being launched at $\sim$ 1000 {\kms} via absorption-line detections of entrained gas from the front and back side of the outflow cone along the inner galaxy sightline to QSO PDS456. 

In this paper we present a more comprehensive survey of the northern Fermi Bubble. We trace the outflowing gas along lines of sight inside and outside the northern Fermi bubble and constrain the radial profile and spatial extent of the nuclear outflow in the Milky Way. The paper is organized as follows. In section 2 we describe the observations and the data reduction. In section 3 we present the UV absorption line spectra and discuss identification and measurement of the outflowing components. In section 4.1 we present the incidence of high velocity absorption around the northern Fermi Bubble. In section 4.2 we present the radial absorption profile and in section 4.3 we present the metallicity of the outflowing gas. In Section 5 we present numerical kinematic models of a nuclear biconical outflow that are motivated by the component structure observed in our spectra. In section 6 we estimate the minimum mass outflow rates and the minimum total gas mass in the Fermi Bubbles. In section 7 we summarize our findings.

\begin{figure*}
\includegraphics[scale=.55]{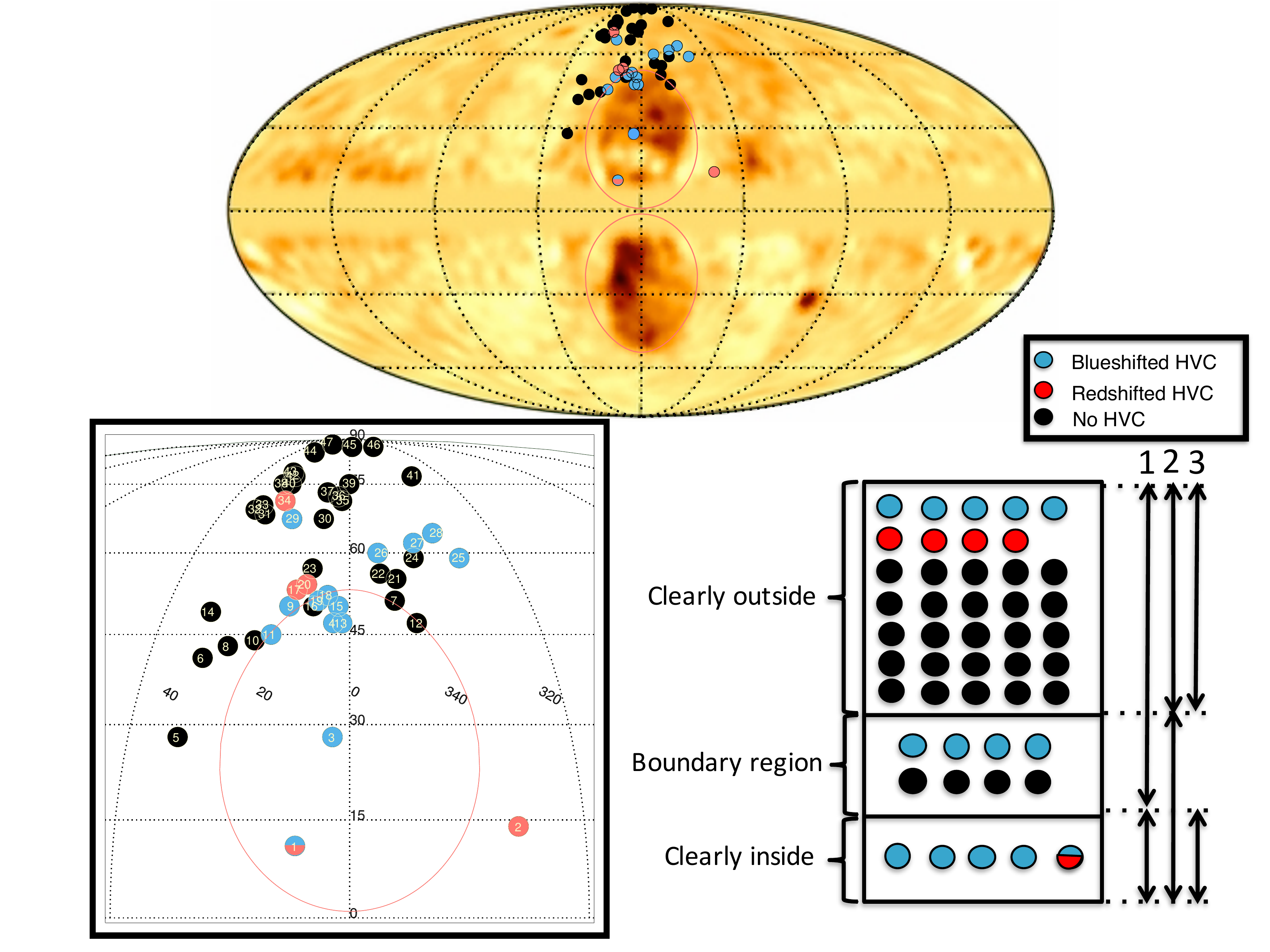}
  \caption{The incidence of high velocity absorption towards the northern Fermi Bubble. Top Panel:-  The all-sky Fermi image of the residual gamma-ray intensity in the 3$-$10 GeV range is shown as the yellow/orange map, in Galactic coordinates centered on the GC  (adapted from \cite{Ackermann2014} and \cite{Fox2015}). The Fermi Bubbles are shown as twin lobes in dark orange at the center. The filled circles mark the position of the lines of sight through the northern Fermi Bubble. The filled black circles represent no HVC detection, the filled blue and red circles represent detected blueshifted and redshifted HVCs, respectively. The circle with both blue and red shadings represents the line of sight with both blueshifted and redshifted HVC. The red contours show the approximate boundary of the Fermi Bubbles. Bottom Left  Panel: Zoomed in map of the top panel. The background sources are marked with their ID numbers from Table \ref{QSO_list}. Bottom Right Panel: Schematic diagram showing the HVC detection statistics in three different regions, using the same symbols, inside and outside the FBs (see text for more details). }
\label{fig:North_map}
\end{figure*}				

\section{Observations and Data Reduction}
 In this section we describe the different observations that are used in this study and how the data are reduced.
\subsection{COS Data}
   
The UV observations for the background quasars were obtained using  the Cosmic Origins Spectrograph (COS; \citealt{Green2012}) on board the Hubble Space Telescope, under the cycle 21 HST Program ID 13448 (PI A. Fox). For five quasars, these observations used G130M/1291 and G160M/1600 grating/central wavelength settings and four FP-POS positions. For one additional quasar we obtained G160M/1600  grating/central wavelength setting observation for which archival G130M observations existed from HST PID 12569 (PI S. Veilleux). For 40 additional background sources we had access to archival G130M+G160M observations from HST/COS (See Table 1). All the archival data were retrieved from the Multi-mission Archive at Space Telescope (MAST) and reduced using \texttt{CALCOS} v3.0 or higher \citep{COSHandbook2016}. We select all available UV bright QSO spectra with a HST/COS spectra that are within  $| l |< $ 35 degrees  and $b>$ 0 degrees. The full target list is shown in Table \ref{QSO_list}. All the targets are selected to lie both inside and outside the northern Fermi bubble (see Figure \ref{fig:North_map}).

All the individual exposures were aligned in velocity space using the centroids of known low-ion interstellar absorption lines and co-added. For a number of lines of sight intergalactic absorption-line systems were also used for wavelength regions without interstellar lines.The final science grade spectra have a signal-to-noise (S/N) near the absorption lines of interest of $\approx$ 12$-$20 (per resolution element), velocity resolution (FWHM) of  $\approx$ 18-20 {\kms} and an absolute velocity scale uncertainty of $\approx $ 5 \kms, and cover the wavelength interval $\approx$ 1150$-$1780 {\AA}. The spectra were continuum normalized around each individual absorption line using a polynomial fit to the continuum. The resulting 1D spectra are binned to Nyquist sampling with three bins per resolution element for display purposes. The analysis and the Voigt-profile fits (described below) were performed on the unbinned 1D spectra.      

In addition, for the 1H1613-097 sightline we performed an orbital night-only reduction of the COS data to remove geocoronal airglow emission. To do this, the spectra were re-extracted using only the time intervals when the Sun altitude as observed by the telescope was less than 20$\degr$. This procedure selects low-background intervals corresponding to the night-side portion of the HST orbit. The spectral re-extraction was conducted with the standard {\tt calcos} pipeline. This process was only conducted for the 1H1613-097 sightline since this is the only sightline in our northern Fermi Bubble sample with a high velocity cloud (HVC) detected in H I 21 cm emission, and therefore where a constraint on [O/H] can be derived (see Section 4.3).

\subsection{GBT spectra}
We obtained deep HI 21 cm data for the four lines of sight (PDS456, 1H1613-097, M5-ZNG1 and MRK1392) inside the Fermi bubble using the Green Bank Telescope (GBT) under program GBT/14B--299, with the goal of detecting the high velocity components in emission. The PDS456 21 cm spectrum is presented in \cite{Fox2015}. Multiple scans of the sightlines were taken using the VEGAS spectrometer in the frequency switching mode. An unconfused velocity range of at least 760 {\kms} about systemic zero velocity at an intrinsic channel spacing of 0.604 {\kms} was obtained while performing the observations by frequency-switching to either 3.6 or 4.0 MHz. The spectra were Hanning smoothed to an effective velocity resolution of 1.2 {\kms}, then calibrated and corrected for stray radiation using the procedure described by \cite{Boothroyd2011}. A fourth order polynomial was fit to emission-free portions of the final average to remove residual instrumental baselines. Such a polynomial fit over a spectrum spanning 760 {\kms} in velocity, will not compromise the measurements of the {\HI} lines, which are only $\approx$ 30 {\kms} in velocity width. The top panels of Figure \ref{fig:line_profile} show the four baseline subtracted 21 cm spectra along each sight-line. The final spectra have typical rms noise values ranging from 8.1 to 12.7 mK temperature brightness per channel. This corresponds to a 1$\sigma$; $N_{HI}$ column density of $\approx \;  1.0 - 1.6 \times 10^{17}$   {\pcm} for a 30 {\kms} wide line.

 \subsection{STIS Data}
For the Halo star M5-ZNG1, the HST/STIS observations were obtained under the HST PID 9410. We refer the reader for observing details to \cite{Zech2008} where this spectrum was published. In short, the observations were taken in 5 orbits in the ACCUM mode with 0.2'' $\times$ 0.2'' aperture. \cite{Zech2008} used the E140M echelle grating to disperse the light onto the far-ultraviolet Multi-Anode Microchannel-Array (MAMA) detector. The spectral resolution is R $\approx$ 45,800 which translates to a velocity resolution (FWHM) of $\approx$ 6.5 \kms.  The STIS data were retrieved from MAST and reduced with the \texttt{CALSTIS} v2.23 pipeline\footnote{\href{http://www.stsci.edu/hst/stis}{http://www.stsci.edu/hst/stis}}. The individual spectral orders were combined into a single spectrum using the IRAF task splice. 

\section{Measurements}
 \subsection{Galactic Center absorber Identifications}
\begin{figure*}
\includegraphics[height=4.5 in, width=2.5in]{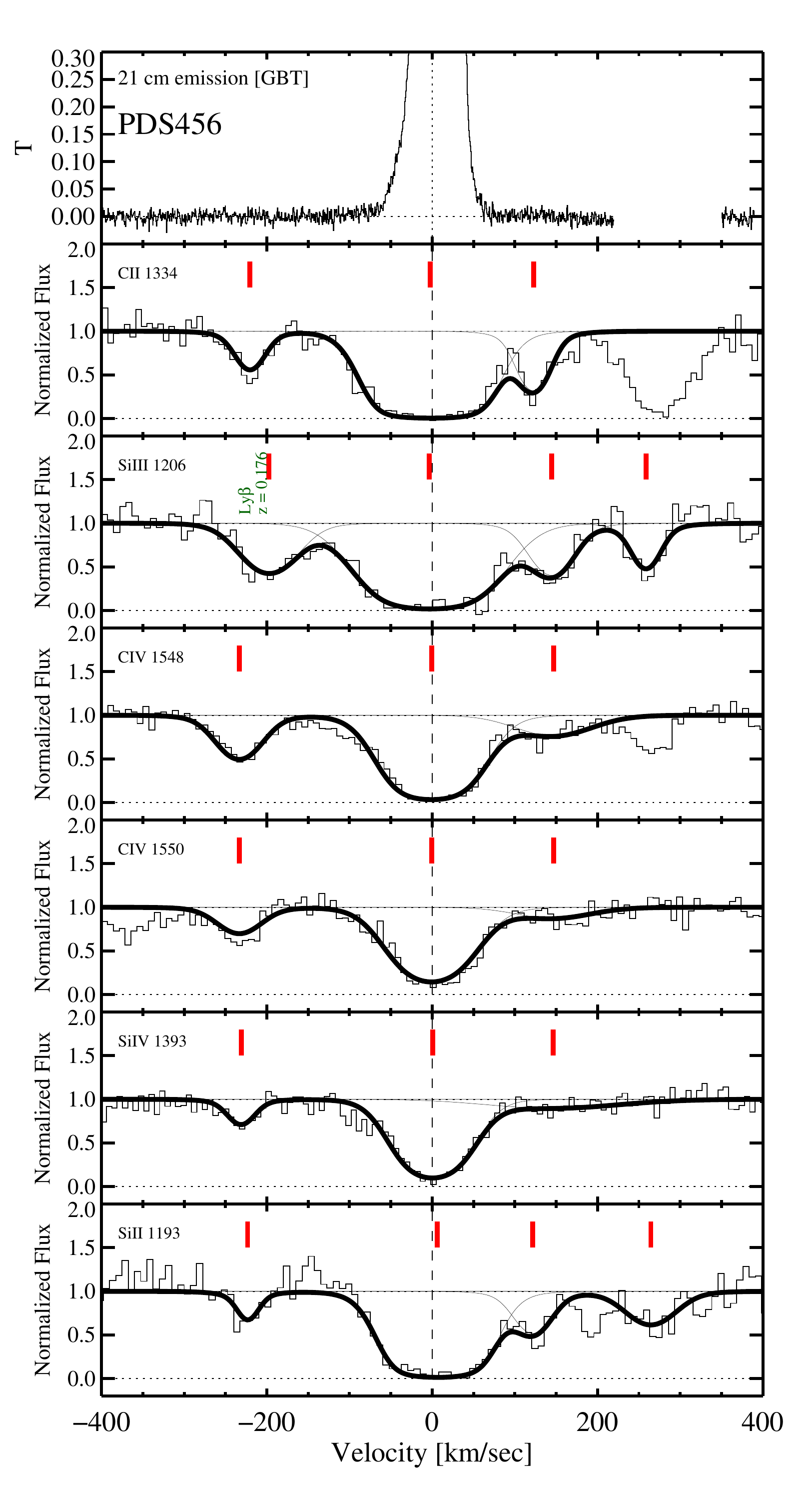}
\includegraphics[height=4.5 in, width=2.5in]{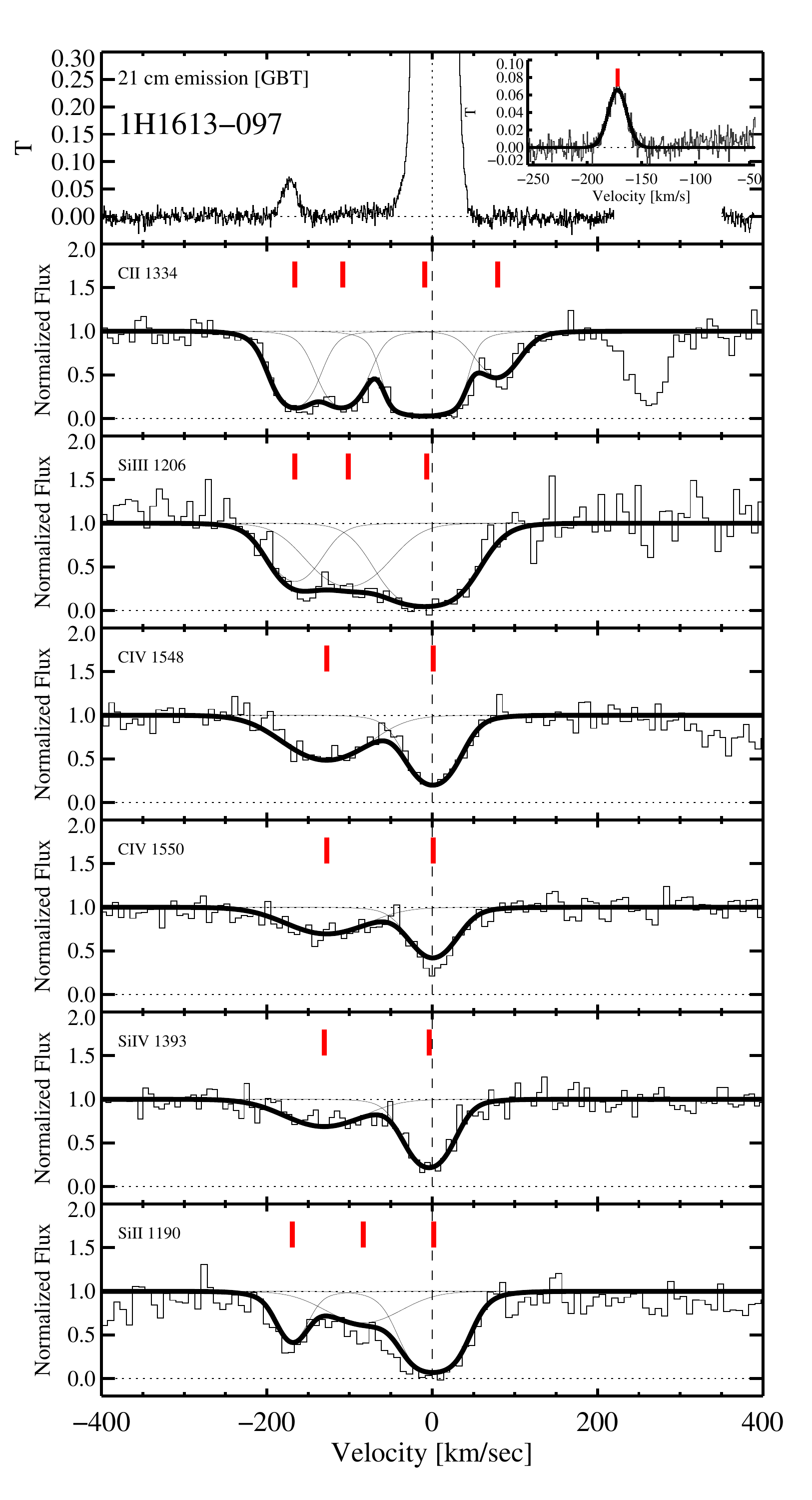}
\includegraphics[height=3.5 in, width=2.25in]{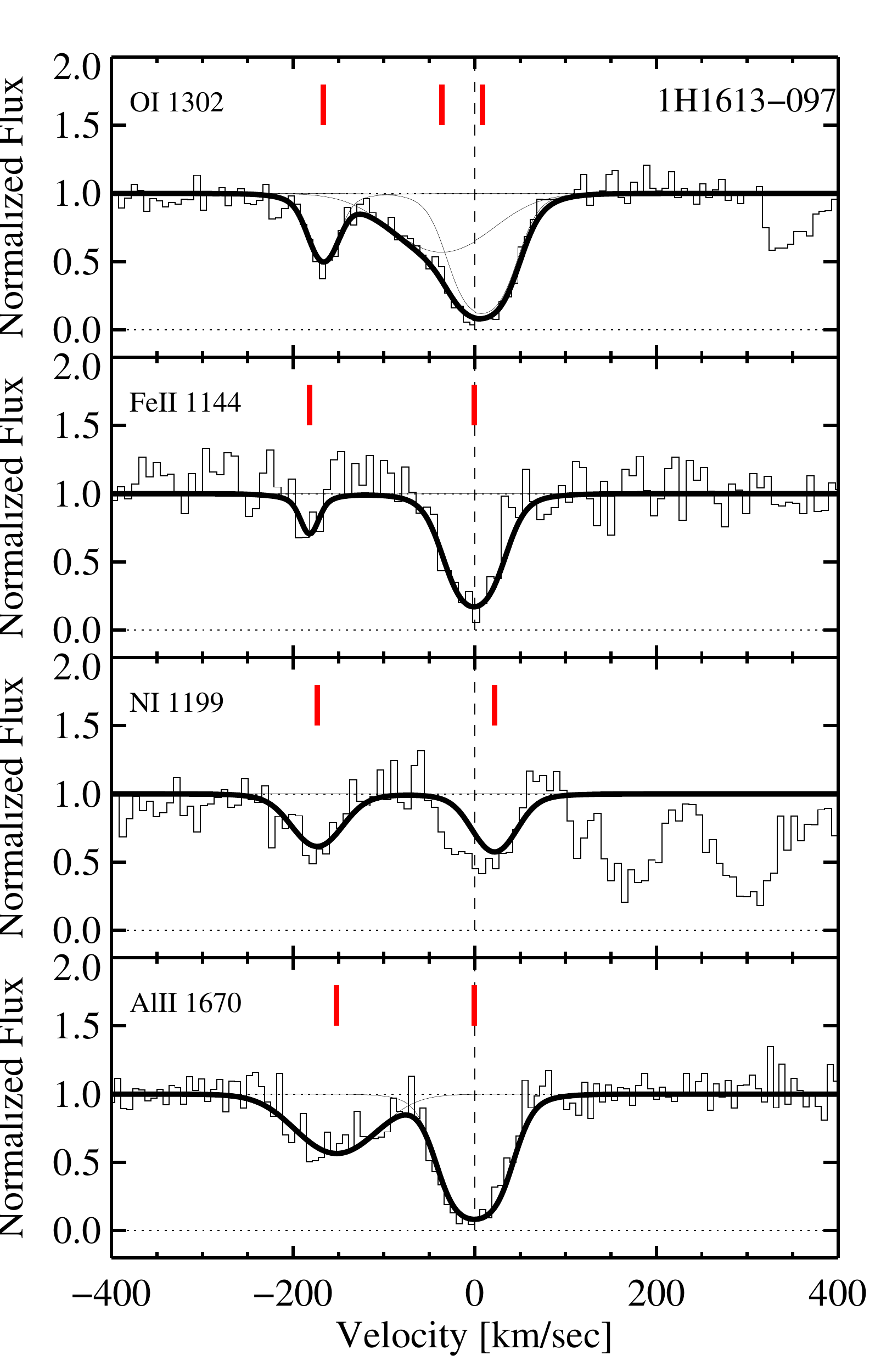}
\includegraphics[height=4.5 in, width=2.5in]{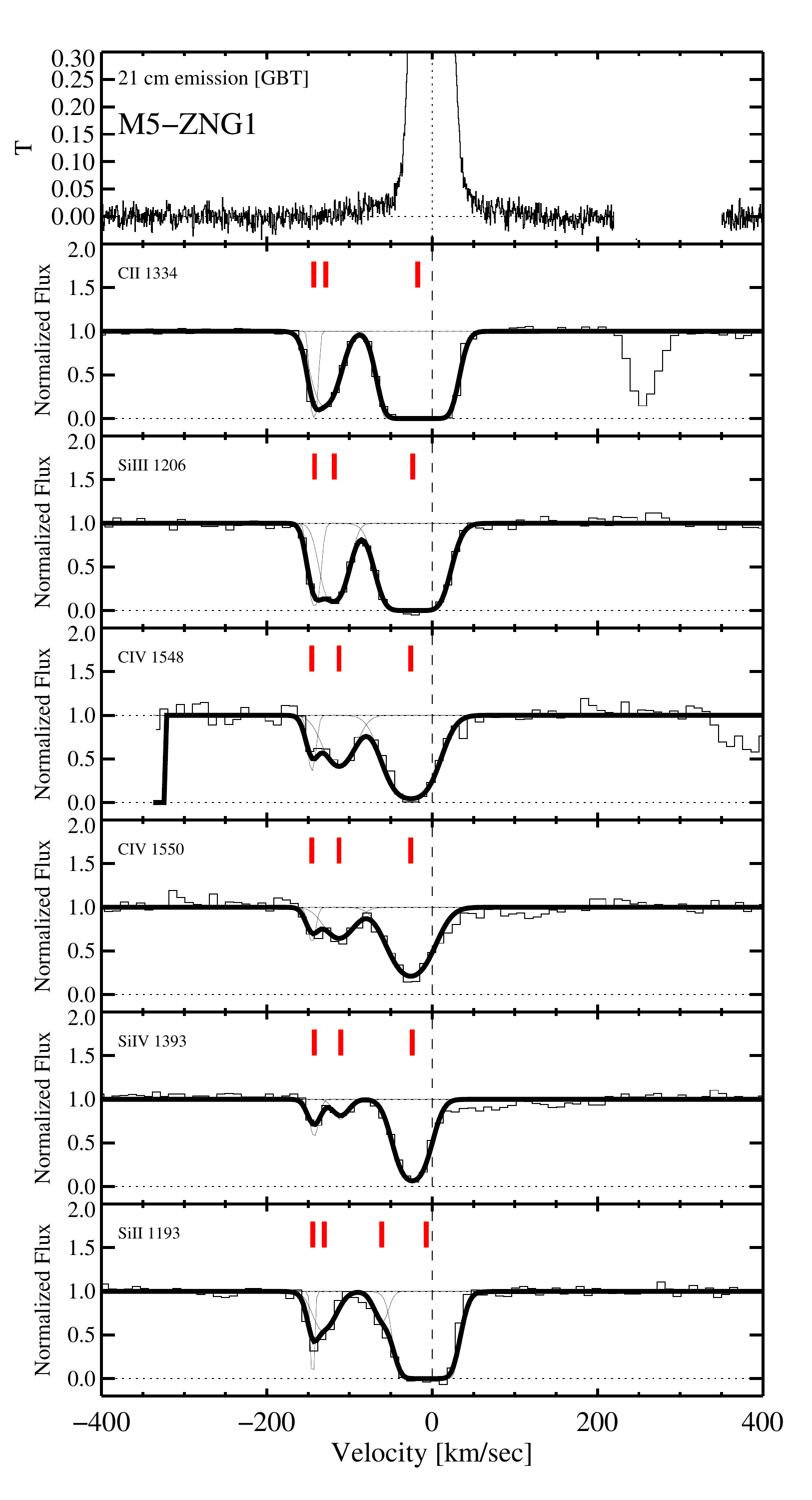}
\includegraphics[height=4.5 in, width=2.5in]{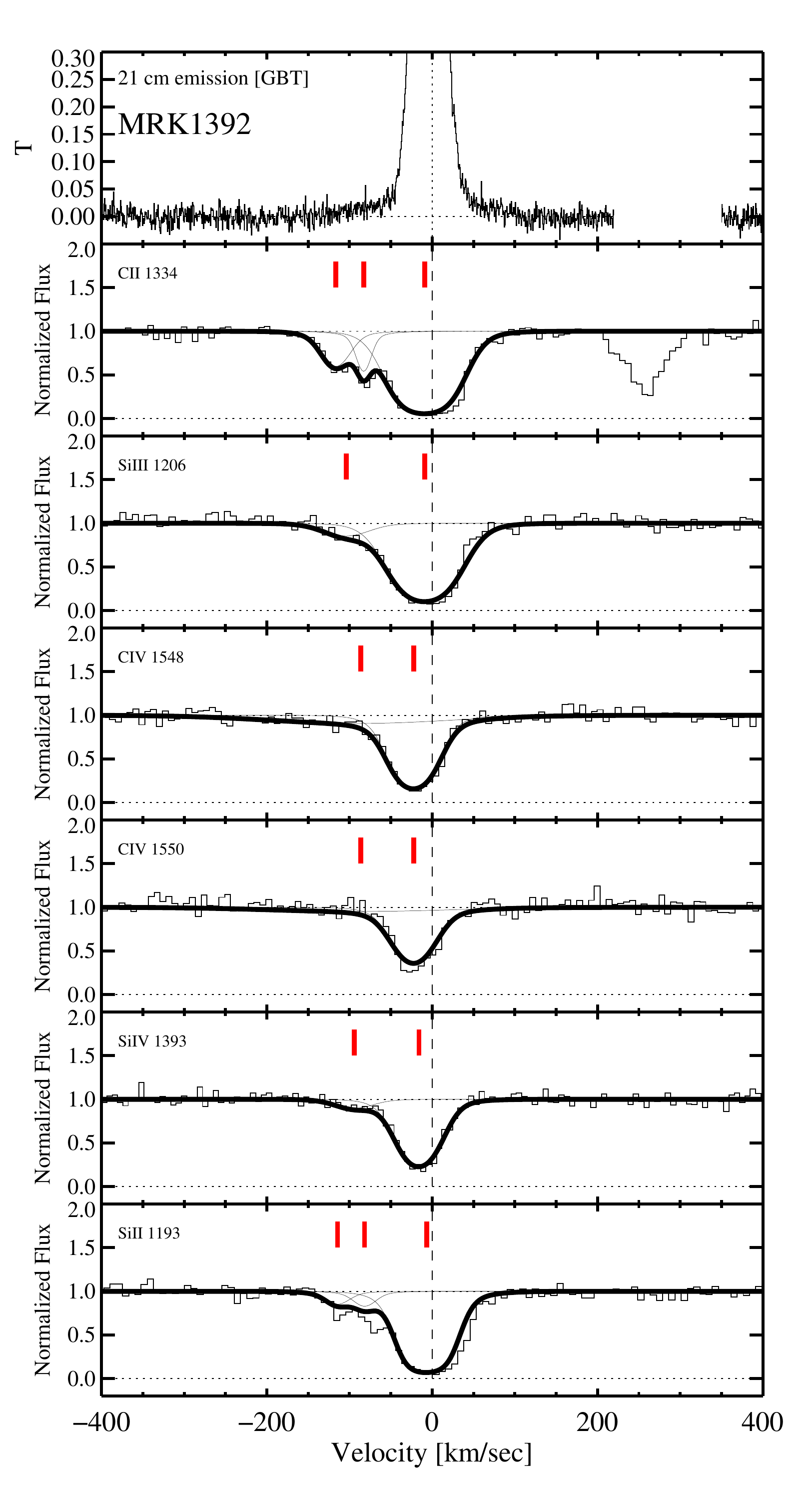}
\includegraphics[height=4.5 in, width=2.5in]{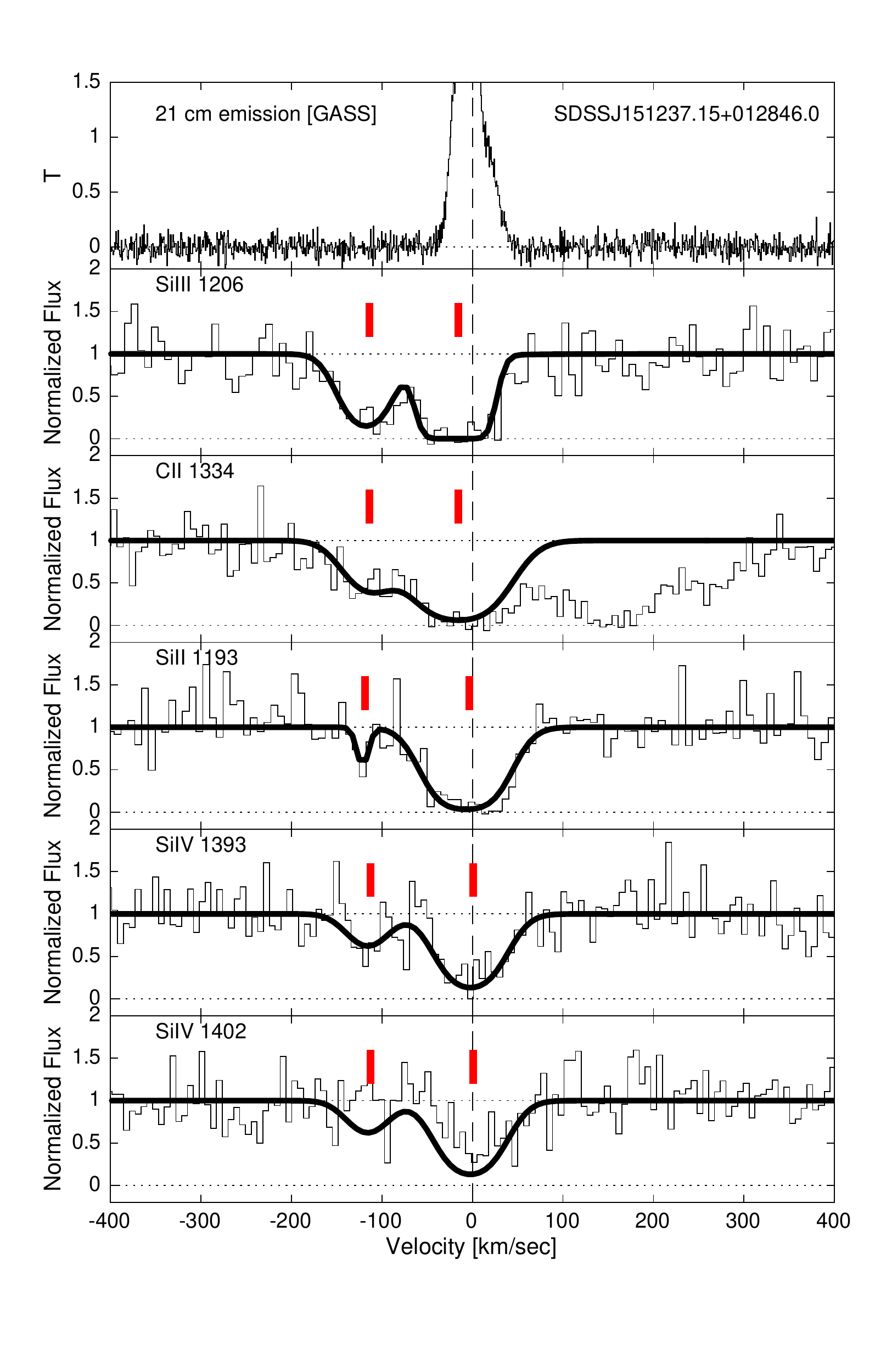}
  \caption{HST UV spectra and 21 cm emission spectra of the five lines of sight within the northern Fermi Bubble. Normalized flux is plotted against the LSR velocity for several UV transitions with their corresponding Voigt profile fits (solid black lines). The vertical red ticks indicate the centroids of individual Voigt profile components. The spectra are re-binned in 3 pixel boxes for presentation. C II*  absorbs at approximately  +264 {\kms} in the rest frame of C II and is responsible for the strong feature near that velocity. For  PDS456, the apparent feature at +260 {\kms} in the spectrum covering {\CIV} at 1548 {\AA} is really the negative velocity feature coming from the {\CIV} 1550 {\AA} transition. We do not show the  GBT spectra between +220 and +350 {\kms} due to base line subtraction issues.}
\label{fig:line_profile}
\end{figure*}

We visually inspect each spectrum to identify any absorption components within $\pm$400 {\kms} of the systemic zero velocity of the Milky Way.  We search for absorption in low-ionization (\CII, \SiII), intermediate-ionization (\SiIII), and high-ionization (\CIV, \SiIV) species in all lines of sight. We classify an absorption system to be a high velocity one if it is detected in multiple species (usually low-ionization lines such as \CII, \SiII, \SiIII, but also in high-ionization lines such as {\CIV} and \SiIV), and if the velocity centroid of that system has $|v_{\rm LSR}|>100$ {\kms}.  We also explore the effect of using a deviation velocity definition of HVCs (see Section 4.1).  We inspect each individual spectrum and identify all detected absorption features associated with the Milky Way, high velocity clouds, higher redshift intervening absorbers and QSO features. We mark the detected intervening higher redshift absorbers in Figures \ref{fig:line_profile} and \ref{fig:appendix}. For each line of sight, we apply a shift to the data to transform the heliocentric velocities provided by COS and STIS to local standard of rest (LSR) velocities as follows:

\begin{equation}
\begin{split}
& \Delta v_{LSR} = v_{LSR} -v_{helio} =\\
&   9 \cos(l) \cos(b) + 12 \sin(l) \cos(b) + 7 \sin(b).
 \end{split}
\end{equation} 
We also transform the LSR velocities to the Galactic standard of rest (GSR) velocities with 
\begin{equation}
v_{GSR} \;=\; v_{LSR} + (254\, \text{kms}^{-1}) \sin(l)  \cos(b), 
\label{eqn_gsr_to_lsr}
\end{equation}
where the rotation velocity $v_R=254\,{\rm km~s}^{-1}$ at the Sun's distance from the Galactic center \citep{Reid2009}. The column densities were determined by independently fitting Voigt profiles to each ion with the \texttt{VPFIT} software \footnote{Available at \href{http://www.ast.cam.ac.uk/~rfc/vpfit.html}{http://www.ast.cam.ac.uk/$\sim$rfc/vpfit.html}} \citep{2014ascl.soft08015C}, using simultaneous fits to all available lines of a given ion.  Note that the line spread function (LSF) of the COS spectrograph is not a Gaussian. For our Voigt profile fit analysis the intrinsic model profiles are convolved with the COS LSF as given at the nearest observed-wavelength grid point in the compilation by \cite{Kriss2011}. For the STIS spectrum, the STIS E140M LSF from the STIS Instrument Handbook\footnote{Available at \href{http://www.stsci.edu/hst/stis}{http://www.stsci.edu/hst/stis}}  was used.

The ionic column densities and Voigt profile fits for the absorption components are reported in Table~\ref{table:Vpfit_measurements}. For each ion, the centroid of the best fit Voigt profile  is used to define the position of each absorption component.

Figure~\ref{fig:line_profile} shows resonant UV absorption line transitions from HST/COS or STIS and {\HI} 21 cm emission spectra from GBT (or GASS for SDSSJ151237.15+012846.0; \citealt{GASS2015}), of the five lines of sight within the northern Fermi Bubble. For each of the absorption line, their corresponding Voigt profile fits are also shown. The velocity centroids of all individual absorption components are shown with a vertical red tick. Below we describe the absorption observed along these five lines of sight, that clearly pass through the northern Fermi bubble in more detail. 

\textbf{PDS 456:}  The absorption components observed along the PDS456 sightline were described in detail in \cite{Fox2015}. To summarize, four absorption components (see top left panel, Figure~\ref{fig:line_profile}) are observed centered at $v_{LSR}$ = $-$235, $-$5, +130, +250 {\kms}, respectively.  We detect low-ionization (\CII , \SiII), intermediate ionization (\SiIII, \AlII), and high-ionization (\CIV, \SiIV, \NV) species, but the relative strength of absorption differs between components. We detect the +250 {\kms} component in the low and intermediate ions only (not in \CIV, \SiIV, and \NV).  The {\SiIII} absorption component at $-$235 {\kms} is blended with a $\rm{Ly -} \beta$ absorption line at $z \,=\,$ 0.176. There might be another -77 {\kms} absorption component seen in weaker S II, P II, Fe II, and C II* absorption.  Figure \ref{fig:appendix2} left panels show additional high velocity blue and redshifted absorption traced by Si II 1260, Si II 1526, Fe II 1144 and Al II1670 ions. In the 21 cm GBT spectrum, we only detect the Milky way component of the {\HI} emission. We do not detect any {\HI} components associated with the high velocity absorption components down to a  $3\sigma$ upper limit $N_{HI} < 3.0 \times 10^{17}$ \pcm.

\textbf{1H1613-097:}  We detect three absorption components (see top middle and top right panels, Figure~\ref{fig:line_profile}) centered at (measured from {\SiIII} 1206 transition) $v_{LSR}$ = $-$164, $-$90, and $-$4 {\kms}, respectively. Some or all of these absorption components are detected in low-ionization (O I, Fe II,  \CII , \SiII), intermediate ionization (\SiIII, \AlII), and high-ionization (\CIV, \SiIV) species, respectively (See Table~\ref{table:Vpfit_measurements} for measurement details). A fourth +83  {\kms} component is also observed in the {\CII} and possibly in Si II 1260 transitions. One {\HI} 21 cm emission line is detected at  $v_{LSR} = -$172 {\kms} along with the Milky Way component, in the GBT spectrum. This {\HI} 21 cm emission component translates to a measured  $\log \;N_{H I} = 18.23 \pm 0.03$. We note that the {\HI} 21 cm emission component is offset from the {\OI} absorption component by 9 {\kms}. After accounting for the COS absolute velocity scale uncertainty of $\approx $ 5 \kms and the Voigt profile fitting error, this offset is quite small ($\approx$ 3 {\kms}) and probably is the result of the COS calibration error or beam smearing effects.  The sub panel at the top right panel of Figure~\ref{fig:line_profile} shows the zoom-in version of this {\HI} 21 cm emission component. There is an emission bump seen at -100 {\kms}, however this emission is not statistically significant to merit a detection. Fitting a Gaussian profile to this excess emission yields a  $\log \;N_{H I} \approx 17.69$  at 2.85$\sigma$ significance. The detected {\OI}, N I, {\FeII} and {\AlII} ions, along with their best fit Voigt profiles are also shown in Figure \ref{fig:line_profile}, top right panel.  Figure~\ref{fig:appendix2} shows additional spectra exhibiting blueshifted high velocity absorption along 1H1613-097 in Si II 1260, Si II 1526, NI 1200c,  and Fe II 1608 transitions.

\textbf{M5-ZNG1:} The absorption components observed along the M5-ZNG1 sightline with HST/STIS spectrum, were described in detail in \cite{Zech2008}. M5-ZNG1 is a halo star 7.5 kpc from the Sun. Three {\SiIII} absorption components (see bottom left panel, Figure~\ref{fig:line_profile}) are seen centered at $v_{LSR}$ = $-$142, $-$118, and $-$24 \kms, respectively. These absorption components are also seen in low-ionization (\CII , \SiII, \AlII), and high-ionization (\CIV, \SiIV, \OVI) species.  We do not detect any blue shifted {\HI} emission component in the GBT spectrum down to a  $3\sigma$ upper limit  of $N_{HI} < 4.8 \times 10^{17}$ {\pcm}, however  \cite{Zech2008} examined a FUSE spectrum to estimate a mean {\HI} column density of $\log N_{HI} = 16.50 \pm 0.06$ using the Lyman series from {\HI} 926 down to {\HI} 918 {\AA}.  \cite{Zech2008} also measured the metallicity of these blueshifted absorption components to be [O/H] = +0.22 $\pm$ 0.10.   Figure~\ref{fig:appendix2} shows additional STIS spectra exhibiting blueshifted high velocity absorption along M5-ZNG1 in Si II 1260, Si II 1526, Al II 1670,  and Fe II 1608 transitions.

\textbf{MRK1392:} Along this line of sight, we detect  absorption components (see bottom middle panel, Figure~\ref{fig:line_profile}), centered at $v_{LSR}$ = $-$117, $-$83, and $-$9 {\kms}, respectively. Blueshifted  absorption components are detected in low-ionization (\CII , \SiII, \AlII~ and \FeII), and high-ionization (\CIV, \SiIV) ions.  For the {\AlII}, {\FeII}, and {\SiIV} transitions, only two absorption components centered at $\approx$ $-$83 and $-9$ {\kms} are detected.  No blueshifted {\HI} emission component is detected in the GBT spectrum down to a  $3\sigma$ upper limit $N_{HI} < 4.8 \times 10^{17}$ \pcm.  Figure~\ref{fig:appendix2} shows additional  spectra exhibiting blueshifted high velocity absorption along MRK1392 in Si II 1260, Si II 1526, Al II 1670,  and Fe II 1144 transitions.

\textbf{SDSSJ151237.15+012846.0:}  The HST/COS spectrum of SDSSJ151237.15+012846.0  covers the G130M grating only. Along this sightline we detect two {\CII}, {\SiII},  {\SiIII}, and {\SiIV} absorption components (see bottom right panel, Figure~\ref{fig:line_profile}), centered at $v_{LSR} \, \approx$  $-$114, and 4 {\kms}, respectively.  Along this line of sight, we do not have any GBT spectrum. We inspect the 21 cm GASS spectrum \citep{GASS2015}, and find no blueshifted {\HI} emission down to a  $3\sigma$ upper limit $N_{HI} < 3 \times 10^{18}$ \pcm.  Figure~\ref{fig:appendix2} top right panel shows the additional Si II 1260 transition exhibiting blueshifted high velocity absorption.

\section{Results}
In this section we report the covering fraction, radial absorption and velocity profiles and metallicity of the high velocity absorption components observed along the northern Fermi Bubble directions.

 \subsection{Incidence of high velocity absorption}    
We measure the observed incidence of high velocity  ($|v_{\rm LSR}|>100$ \kms) absorption components inside and outside the northern FB. To quantify if a line of sight passes through the FBs, we have to define their boundaries. These boundaries are not well defined because of noise in the gamma-ray data. We therefore visually inspect the gamma-ray map of the FBs and approximately define the northern FB to be confined within a circle of radius 26 degrees centered at $l$ = 0, and $b$ = 27 degrees. 

This boundary is shown with the red contours on Figure \ref{fig:North_map}, and allows us to divide the sightlines into three categories: through the Bubble (5 sightlines), outside the Bubble (34 sightlines), and interface region (8 sightlines). The interface sightlines straddle the boundary region of the northern FB, and might probe the kinematics of cool and warm gas at the edge of the northern FB. We will compute the incidence of HVCs in three cases:  Case 1, where the boundary absorbers are treated as being outside the FB. Case 2, where they are treated as inside the FB. Case 3,  where they are omitted from the calculation, to explore the sensitivity of the HVC covering fractions to the definition of the boundary. The bottom right panel of  Figure \ref{fig:North_map} shows a schematic diagram of the HVC detection statistics in these three regions. The three cases are marked as 1, 2 and 3 respectively.

\textbf{Case 1:} We first assume that all the eight interface lines of sight are passing outside the Fermi Bubbles (see Figure \ref{fig:North_map}, bottom right panel). In this case, all five lines of sight that pass through the northern FB exhibit blueshifted high velocity absorption, and nine out of  42 lines of sight outside the northern FB exhibit blueshifted high velocity absorption (see Figure \ref{fig:North_map}).  To quantify this, we use the Wilson score interval to estimate the underlying binomial hit rates. The 5 out of 5 lines of sight with blueshifted  high velocity absorption inside the northern FB yield an incidence rate of 92$\pm$ 8\%, and 9 out of 42 sightlines with blueshifted  high velocity absorption outside the FB yield an incidence rate of 22$\pm$ 6\% (9/42). These are shown in Figure \ref{fig:appendix} and the detections are tabulated in Table \ref{appendix_table}.

Inside the northern FB,  only one line of sight (PDS456), exhibits both blueshifted and redshifted high velocity absorption components, whereas none of the lines of sight out of the 42 outside the FB exhibits both blueshifted and redshifted high velocity absorption. Inside the bubble, the presence of both blueshifted and redshifted HVC components, which we previously argued in \cite{Fox2015} may be as the signature of the biconical outflow, is only seen in PDS456. However, the lack of redshifted absorption components at higher latitudes can be understood in the context of an outflow model as a geometrical effect: at higher latitudes, the nearside of the outflow cone is much closer to the observer than the far side, and the lines of sight pass through only the nearside of the outflow cone. Hence we only see the blueshifted HVCs (and not the redshifted HVCs) for the three lines of sight inside the FB at high latitudes. The low-latitude lines of sight,  pass through both sides of the outflow cone at similar z-distances, resulting in our observing both blueshifted and redshifted high velocity absorption components.

At $b <30$ degrees two out of four lines of sight exhibit redshifted HVC absorption with an incidence rate of 50 $\pm$ 22 \%. One  redshifted absorber outside the FB (along QSO1503-4140) and one inside the bubble (along PDS 456) are detected. It should be noted that the effect of galactic rotation must be accounted for while analyzing the two low $b$ directions outside the FB as foreground gas co-rotating with the Milky Way disk can produce absorption at a range of observed velocities. Assuming cylindrical co-rotation, a simple model of Galactic rotation can be used to predict the maximal allowed velocities for a given latitude and longitude (e.g.,\citealt{Wakker1991}). We find that the none of the extreme blueshifted or redshifted velocities are consistent with co-rotating foreground gas. We would require new data to better quantify the statistics of low $b$ redshifted HVCs.  For all $b$, the incidence of redshifted HVCs inside the northern FB is 25 $\pm$ 17\% (1/5), and outside the northern FB is 10 $\pm$ 5\% (4/42). The incidence of any (blueshifted or redshifted) high velocity absorption inside the FB is 92$\pm$ 8\% (5/5), and outside the FB is 31 $\pm$ 7\% (13/42). 

\textbf{Case 2 :} We now extend the definition of boundary of the Fermi Bubbles to explore the effect of assuming that the eight boundary lines of sight are inside the northern FB (see Figure \ref{fig:North_map}, bottom right panel). In this case, 9 out of 13 lines of sight inside the northern FB exhibit  blueshifted high velocity absorption and yield an incidence rate of 68$\pm$ 12\%, and 5 out of 34 sightlines with blueshifted high velocity absorption outside the northern FB show an incidence rate of 16 $\pm$ 6\%. For redshifted HVCs, the incidence rate inside the northern FB is 11 $\pm$ 8\% (1/13),  and outside the northern FB is 13 $\pm$ 6\% (4/34). The incidence rate of any (blueshifted or redshifted) high velocity absorption inside the northern FB is  68$\pm$ 12\% (9/13), and outside the northern FB is 27 $\pm$ 7\% (9/34).

\textbf{Case 3 :} Finally, as all the eight lines of sight that pass through the boundary of the northern FB might contain complex kinematics due to shocks from the nuclear outflow terminating in those regions;  we exclude these eight sightlines while calculating covering fractions  (see Figure \ref{fig:North_map}, bottom right panel). The  covering fraction of blueshifted  HVC absorption outside the northern Fermi Bubble becomes  16 $\pm$ 6\% (5/34), the covering fraction of redshifted HVCs outside the northern FB becomes 13 $\pm$ 6\% (4/34),  and the covering fraction of all HVCs outside the northern Fermi Bubble is 27\% (9/34) $\pm$ 7\%. Inside the northern FB, the covering fraction of blueshifted and any absorbers is 92$\pm$ 8\% (5/5), and the incidence of redshifted HVC is  25 $\pm$ 17\% (1/5).

\textit{In all three cases we see that the rate of incidence of blueshifted HVCs inside the northern FB is always higher than that outside the northern FB.} In all three cases, we perform the adjusted Chi-squared test with the Yate's correction for continuity on the blueshifted high velocity absorbers. The P values for cases 1, 2 and 3 are 0.0018, 9.6784e-04 and, 4.1584e-04, respectively. These P values suggest that we can rule out the null hypothesis that the distribution of blueshifted high velocity absorbers inside and outside the northern Fermi bubble are the same at more than 99.8\% confidence level. We further perform this test for any (blueshifed or redshifted) high velocity absorption and find that the P values for cases 1, 2 and 3 are 0.011, 0.018 and, 0.0068 respectively.  These P values indicate that we can rule out the null hypothesis that the distribution of any high velocity absorbers inside and outside the Fermi bubble are the same more than 98.2\% confidence level. The significantly higher incidence inside and lower incidence outside the northern FB suggests that the FBs contains a  reservoir of entrained cool gas that is confined to the same physical regions as traced by gamma ray emission maps.  Overall, the incidence of redshifted HVCs are not distinguishable inside or outside the FB. But we have very few redshifted HVCs detected in this survey. Particularly at lower galactic $b$, new data would be required to improve on the statistics of redshifted HVCs.

We further computed the incidence of high velocity absorption using the deviation velocity method to identify a high velocity absorber \citep{1997ARA&A..35..217W}.  We quantify any absorber as a high velocity  absorber if  $|v_{dev}| \geq $ 80 {\kms}.  Using this method, we find that all 5 lines of sight inside the northern Fermi Bubble are blueshifted HVCs with an incidence rate of 92$\pm$ 8\%.  If we include the interface sightlines to be outside the FBs, 13 out of 42 sightlines outside the northern FB exhibit blueshifted HVCs with an incidence rate of 31$\pm$ 7\%. Out of the 8 interface sightlines 6 exhibit blueshifted HVCs, making the incidence of blueshifted HVC in the interface region alone to be 72 $\pm$ 15\%(6/8). If we do not count the interface sightlines as being outside the FB, 7 out of 34 sightlines show a deviation velocity HVC outside the FBs ( 21 $\pm$ 7\%). With both methods of identifying a HVC, we see an excess incidence of blueshifted HVCs inside the northern FB compared to outside of it. For redshifted HVCs, incidence absorbers inside the northern FB is 25 $\pm$ 17\% (1/5), and outside the northern FB is 10 $\pm$ 5\% (4/42).   Using the deviation velocity method, the incidence of any (blueshifted or redshifted) high velocity absorption inside the FB is 92$\pm$ 8\% (5/5), and outside the FB is 31 $\pm$ 7\% (13/42). 

Our GC sightlines pass above the Scutum-Centaurus spiral arm, which may produce its own star-formation-driven outflows. These outflows represent a foreground signal to be removed when searching for the absorption components due to the GC outflow. Spiral-arm outflows are known to produce detectable signatures in UV absorption in sightlines towards background targets \citep{Tripp1993, Fox2003, Lehner2011}. These foregrounds are strongest at low latitude, and are not modeled in depth here. However, the clear difference we measure in the covering fractions for inside-the-Bubble sightlines vs outside-the-Bubble sightlines supports the interpretation that the HVC components trace the nuclear wind and not a foreground.

\begin{figure*}
\centering
\includegraphics[height=7.cm,width=8.75cm]{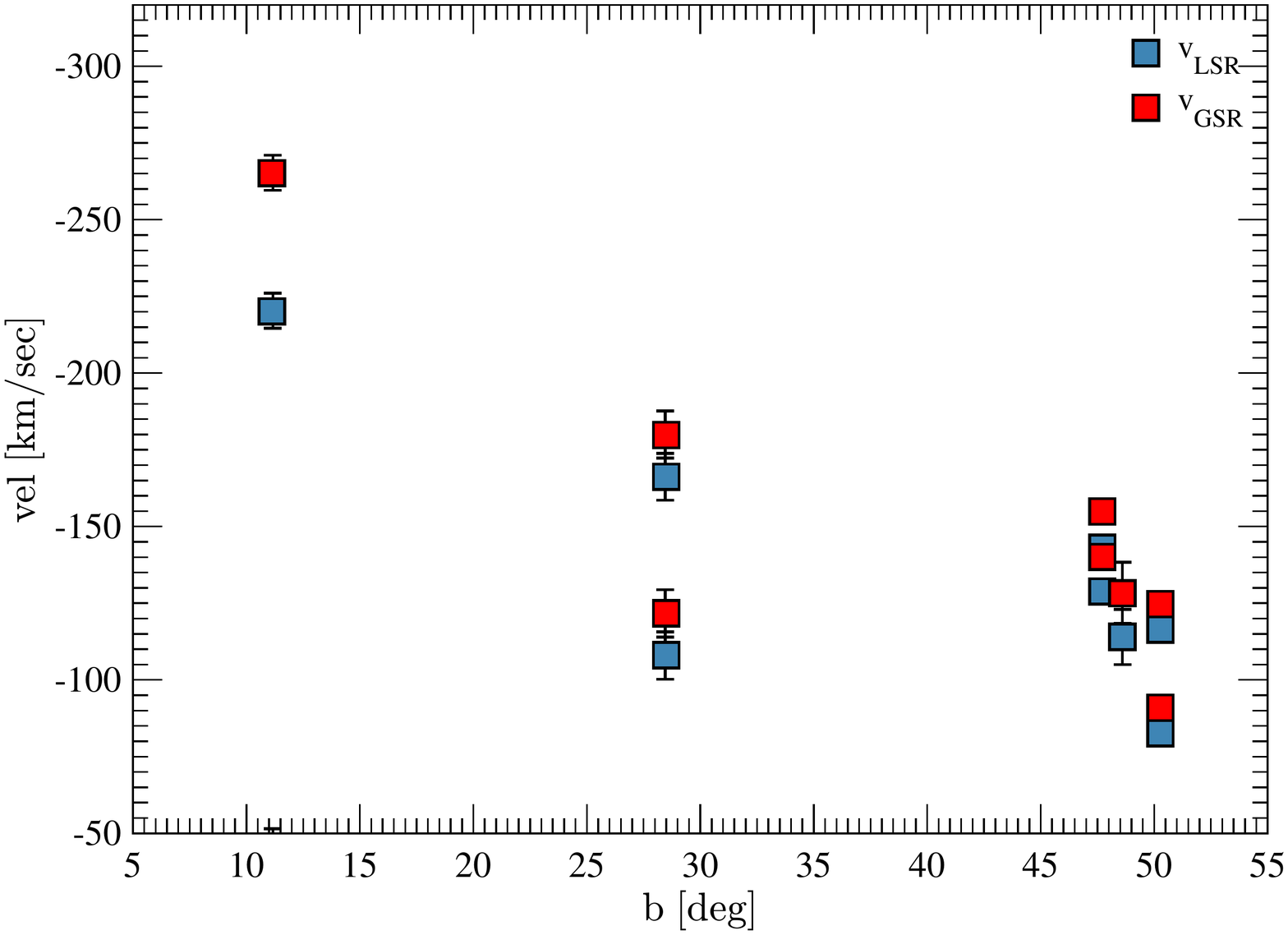}
\includegraphics[height=7.cm,width=8.75cm]{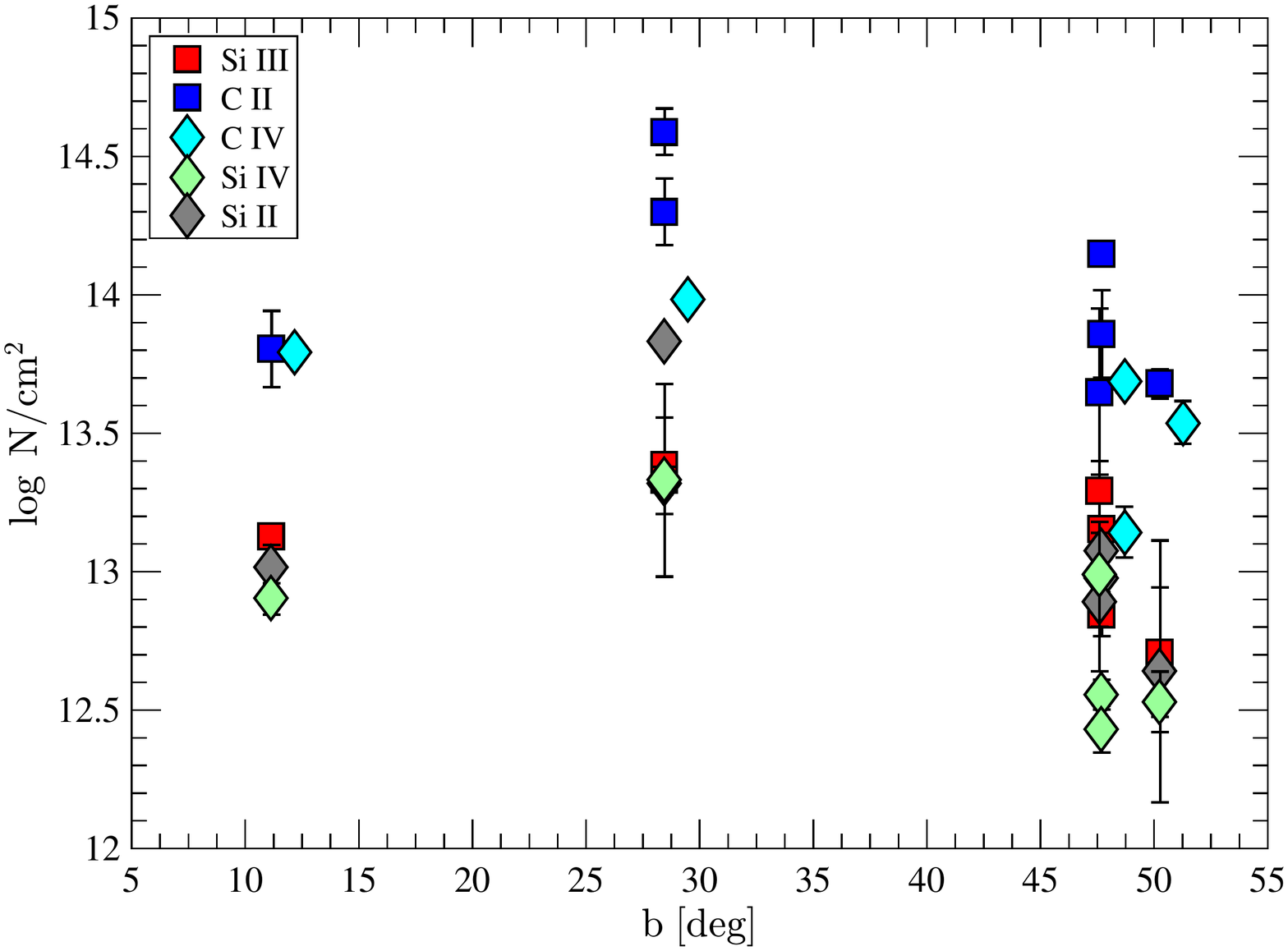}
  \caption{Left Panel: Radial velocity profile of the blueshifted {\CII} high velocity absorption inside the northern Fermi Bubble (sightlines located outside the northern Fermi Bubble are not shown), as a function of Galactic latitude. The blueshifted outflow velocities decrease with increasing higher Galactic latitudes both in LSR velocity (blue squares) and GSR velocity (red squares). The error bars are the uncertainty on the velocity centroid in the Voigt profile fits. For 1H1613-097, M5-ZNG1 and MRK1392 we resolve the blueshifted high velocity absorption into two individual absorption components. Right Panel:  Radial absorption profile of the blueshifted high velocity absorption inside the Fermi Bubble as a function of Galactic latitude. The column densities are Voigt profile fitted column densities for each species. The C IV column density profile is plotted with a 1 degree offset along the x-axis for presentation.}
\label{fig:radial_profile}
\end{figure*}		

\subsection{Radial Dependence of Absorption}
To show the gas kinematics and absorption strength of the gas inside the Fermi Bubbles, we study the variation in kinematics  and column density of the blueshifted high velocity absorption, as a function of Galactic latitude for the five lines of sight that passes through the northern Fermi Bubble. These five lines of sight are roughly at similar galactic longitudes, which allows us to study the variation of gas kinematics at different latitudes directly above the GC. We use {\CII} Voigt profile velocity centroids to quantify the kinematics of the blueshifted high velocity absorption. This line was chosen because it is a strong low-ion transition detected in each sightline passing through the FB.

Figure~\ref{fig:radial_profile}, left panel shows the radial velocity profile of the blueshifted high velocity absorption inside the Fermi Bubble as a function of Galactic latitude. Both the GSR (red square) and LSR (blue square) velocities are shown. For three lines of sight (1H1613-097, M5-ZNG1, and MRK1392), we resolve the blueshifted {\CII} high velocity absorption into two individual absorption components. Such multiple blueshifted absorption may represent gas  that is entrained within the bipolar outflow cone. In other words, we are maybe not only seeing gas at the edge of the outflow cone, but also some absorption that is inside the outflow cone. We observe a trend of decreasing blueshifted outflow velocity with increasing Galactic latitude and radial distance from the Galactic center. The observed velocity changes from $v_{GSR}$ = $-$265 {\kms} at  $b \sim$11$^{\circ}$ to  $v_{GSR}$ = $-$91 {\kms} at $b \sim$ 50$^{\circ}$.

Figure~\ref{fig:radial_profile}, right panel shows the radial absorption profile of the same blueshifted high velocity absorption inside the Fermi Bubble, as a function of Galactic latitude. We show the observed column densities of {\SiII}, {\SiIII}, {\SiIV}, {\CII} and {\CIV} transitions, respectively. We do not see any radial trend of absorption column density with galactic latitude. However,  as we probe  $b > 45^{\circ}$, marginally weaker {\SiIII}, {\SiIV}, and {\SiII} absorption components are detected which are not seen in more close in lines of sights. These findings show that the entrained gas in the Fermi Bubble is seen at least out to a latitude of $\approx$ 50$^{\circ}$ from the Galactic Center beyond the covering fraction of blueshifted high velocity gas  rapidly falls off.

\begin{figure}[!t]
\includegraphics[width=8.5cm, height=9.5 cm]{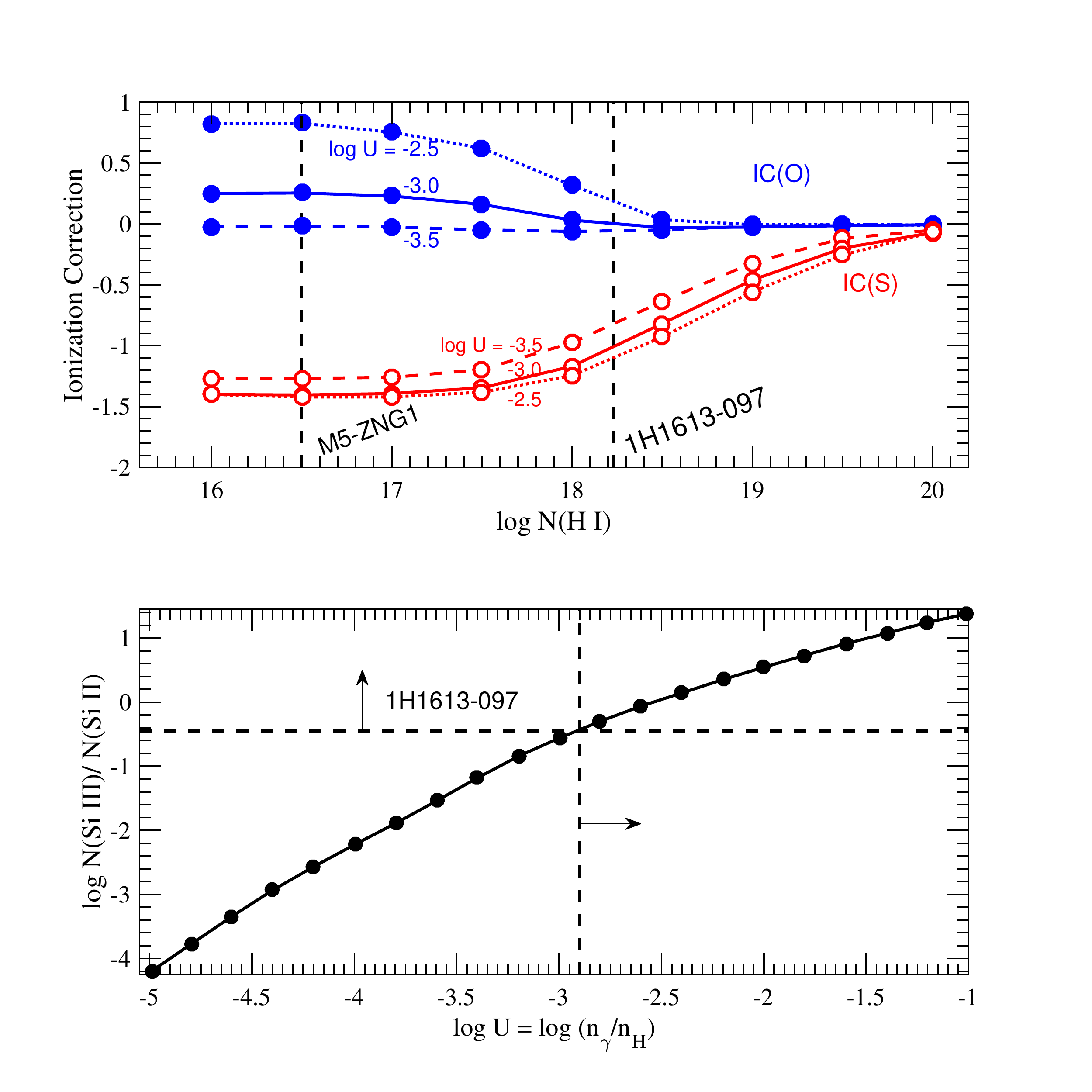} 
\caption{Upper panel: Ionization corrections IC(O) and IC(S) against \HI\ column density for a 
uniform-density photoionized cloud.
The values of $N$(H I) appropriate for the HVCs towards 1H1613-097 and M5-ZNG1 are shown 
with dashed vertical lines. These corrections are used to convert [\OI/\HI] into [O/H]. 
Lower panel: Dependence of the ion ratio $N$(\SiIII)/$N$(\SiII) on log\,$U$ for the case of the 
HVC at $-$172 \kms\ towards 1H1613-097. The measured value of the ratio is used to constrain log\,$U$,
which in turn is used to constrain IC(O).}
\label{fig:cloudy}
\end{figure}

\subsection{Metallicity of the outflowing gas along 1H1613-097}

To constrain the chemical abundances in the Galactic Center HVCs, we analyzed the 
\OI/\HI\ ratio in the component at $-$172 {\kms} towards 1H1613-097
and in the $-$143 and $-$125 {\kms} components towards M5-ZNG1.
These three clouds were chosen because they are the only HVCs in our northern 
Fermi Bubble sample where both the {\OI}  and \HI\ column densities are securely measured.
\OI/\HI\ provides a good metallicity indicator, since O is relatively mildly depleted onto 
dust grains \citep{Cartledge2008,Jenkins2009}, and charge-exchange reactions tie the two species
together \citep{FS71}. However, an ionization correction (IC) may apply if the gas is optically 
thin (e.g. \citealt{Viegas1995}), defined such that
\begin{equation}
       \mathrm{ [O/H]=[O I/H I] + IC(O)}.
\end{equation}

\subsubsection{Methodology of Ionization Modeling}
We used the photoionization code {\it Cloudy} \citep{Ferland2013} to investigate the 
magnitude of the possible IC, using the following steps:

(i) We constructed a grid of {\it Cloudy} models at values of log $N$(\HI) between
16 and 20 in 0.5\,dex intervals, using an ionization parameter log\,$U$=$-$3.0, 
where $U=n_\gamma/n_{\rm H}$, the ratio of the ionizing photon density to the gas density.
We adopt the position-dependent (3D) combined Galactic and extragalactic radiation 
field presented in \cite{Fox2014}, based on \cite{BH1999} and \cite{Fox2005}, 
taken at a distance 10 kpc along the 1H1613-097 sightline. The gas is assumed
to be at uniform density.
In principle, the value of log\,$U$ in an HVC can be derived from observations
of the \SiIII/\SiII\ column-density ratio. However, in the HVC towards 1H1613-097, 
\SiIII\ 1206 appears saturated and so only a lower limit on the ratio, 
$N$(\SiIII)/$N$(\SiII)$\ga-0.45$, can be derived. Fortunately, the ratio is 
unlikely to be much higher than this limit as the {\SiIII} line is not strongly saturated
(Figure \ref{fig:line_profile}, top center panels).

(ii) We use the results of the {\it Cloudy} model to calculate IC(O) and IC(S) at 
each value of log $N$(\HI), producing the curve shown in Figure \ref{fig:cloudy}. 

(iii) We repeat steps (i) and (ii) with log\,$U$ values of $-$2.5 and $-$3.5, to
investigate the sensitivity of IC to the choice of $U$. 

\subsubsection{Results of Ionization Modeling}
The upper panels of Figure \ref{fig:cloudy} shows the run of IC(O) against $N$(\HI). We also include IC(S)
for convenience, derived in an analogous way for \SII\ observations.
The lower panels show the dependence of the ion ratio $N$(\SiIII)/$N$(\SiII) on log\,$U$ 
for the case of the HVC at $-$172 \kms\ towards 1H1613-097.
The model implies log\,$U\ga$-2.9. Although this is formally a limit, it is close to 
values derived for many other Galactic HVCs
\citep{Collins2005, Richter2009, Tripp2012, Fox2014, Fox2016}, and since the saturation in 
\SiIII\ is mild the actual value is unlikely to be much higher.
This constraint on log\,$U$ translates to a constraint IC(O)$\ga$+0.1. Thus the measured oxygen abundance of [\OI/\HI]=$-$0.64 in the HVC needs to be corrected upwards by $\ga$0.1\,dex.

We use the stray radiation correction procedure described in \cite{Boothroyd2011} on the data to account for the structure of the GBT beam at 21 cm, and this procedure should remove any radiation originating outside the main beam to the maximum extent possible. However, since the \HI\ measurement is derived from radio observations using a finite beam, and the UV observations are derived from effectively infinitesimal beams, a beam-smearing error of $\sim$ 0.15 dex must also be taken into account to account for potential small-scale structure in the beam  
\citep{2001ApJS..136..537W}. Therefore the ionization-corrected oxygen abundance in this HVC is 
[O/H]$\ga-0.54\pm0.15$. This is lower than expected for material recently ejected from the GC,
but it is a lower limit, so the true value could be higher. Low metallicity HVCs associated with the GC region have been reported before. 
\cite{Keeney2006}, studying a sightline (PKS 2005-489) passing through 
the high-latitude southern GC region, also reported HVCs with 
$\gtrsim$ 10 $-$ 20\% solar metallicity.

In the two HVCs towards M5-ZNG1, for which \cite{Zech2008} report [O/H]=+0.22$\pm$0.10 \footnote{  $ \mathrm{[O/H] = \log (O/H) - log(O/H)_{\odot}}$.  Note that for $\log(O/H)_{\odot}$ we adopt the \cite{Asplund09} value of -3.31.  However,  \cite{Zech2008} used the  \cite{Asplund2005} value of -3.34.  A correction of 0.03 dex is needed for fair comparison between the two measurements. },
the calculated IC is larger, IC(O)=+0.25, for an assumed log\,$U$=$-$3.0,
because of the significantly lower \HI\ column density, 
log $N$(\HI)=16.50$\pm$0.06 \citep{Zech2008}. However, in this low-column-density regime 
the IC is highly uncertain and subject to charge-exchange reactions, 
so we quote a conservative ionization-corrected [O/H]$>$+0.22 for this HVC.
Therefore the M5-ZNG1 HVC shows a considerably higher value of
O/H than the 1H1613-097 HVC, by $>$0.88\,dex. 
There are several plausible reasons for this difference.
Most notably, the distance to M5 (a globular cluster at 7.5 kpc;  \citealt{Harris1996}) may
place it \emph{in front of} the Fermi Bubble; this would be the case if the radial (line-of-sight)
extent of the Fermi Bubble was the same size as its tangential extent on the sky. In this case,
the blueshifted HVCs in this direction are foreground object unrelated to the GC or the 
Fermi Bubbles. It is also possible that the M5-ZNG1 HVCs are photospheric 
in origin (M5-ZNG1 is a post-AGB star); while \cite{Zech2008} disfavor this idea based on path 
length arguments, the HVCs could still be subject to unusual ionization conditions. 
Further measurements of chemical abundances of HVCs in the GC region
are needed, particularly when derived in combination with ionization corrections.

We do not present a determination of [S/H] in the 1H1613-097 HVC from the \ion{S}{2} 1250, 1253, 1259 triplet 
because the \ion{S}{2} 1259 line appears to be contaminated at the HVC velocity and the 1250 and 1253 lines 
show no significant detection. However, we still show the behavior of IC(S) with $N$(\HI) on Figure 
\ref{fig:cloudy}, since this may be useful for abundance studies in 
other HVCs, where \ion{S}{2} is reliably detected.

\section{Modeling the absorption}

We interpret absorption features in terms of simple models inspired by the observations and theories described earlier.  Our goal is to develop the transformations from three-dimensional velocity vectors at different locations to the (scalar) radial velocities that we can measure.  We start with the expected behavior of outflowing material moving at a velocity $v$ along a trajectory directly away from the Galactic center into the halo.   We assume that there is no coupling of this gas to Galactic rotation, which seems to be supported by observations of compact neutral hydrogen clouds above and below the Galactic plane \citep{McClure-Griffiths2013}. These models are mathematically identical to the simple models first described in \cite{Fox2015} which in turn were based on the Mg II outflow models of \cite{Bordoloi2012a}.

In the simplest picture, the motions are along the edges of nested cones, all of which have their vertices at the Galactic center and axes perpendicular to the plane of the Galaxy.   In a refinement of this picture, we acknowledge that perhaps the gas is not ejected from just the nucleus of our Galaxy, but instead could originate from inside a small, circular zone in the plane of the Galaxy that is centered on the Galactic center.  This picture is consistent with the proposals that the Fermi Bubble outflows are generated by a central region of our Galaxy that has rapid star formation  \citep{Bland-Hawthorn2003, Carretti2013, Crocker2015, Lacki2014}.   
For models that favor Sgr~A* as the origin \citep{Zubovas2011, Guo2012, Bland-Hawthorn2013, Ruszkowski2014, Mou2014}, one could envision that an initial spherical outflow is shaped into a conical one by resistance from the static gaseous layer in the Galactic plane.

\subsection{Conical Outflow from the Galactic Center}
We define a coordinate system centered on the Sun that has an $x$-axis that points towards the Galactic center, a $y$-axis towards the Galactic coordinates $(\ell,~b)=(90\arcdeg ,~0\arcdeg)$ and the $z$-axis towards $b=90\arcdeg$.  We then initially imagine the presence of two vectors that start at the location of the Sun and point towards the Galactic center: one of them that we call $\textit{\textbf{v}}_1$  has a length $R=8.4\,$kpc equal to the distance from the Sun to the Galactic center \citep{Reid2009}, and the other, called $\textit{\textbf{v}}_3$, has a length $\mathcal{L}$, which can be either greater or less than $R$.  If we now rotate $\textit{\textbf{v}}_3$ about the $y$-axis by a Galactic latitude angle $b$ and follow this with a rotation about the $z$-axis by a Galactic longitude angle $\ell$, we then have transformed $\textit{\textbf{v}}_3$ so that it that ends at some location on a conical surface that has its vertex at the Galactic center, where

\begin{equation}\label{v3_noz}
\textit{\textbf{v}}_3={\mathbf R}_z(\ell){\mathbf 
R}_y(b)\left(\begin{array}{c}\mathcal{L}\\ 0\\ 0\end{array}\right)=\mathcal{L} 
\left(\begin{array}{c}\cos \ell\cos b\\ \sin \ell\cos b \\ \sin 
b\end{array}\right)~.
\end{equation}

To find the distance $r$ from the vertex of the cone to the end of 
$\textit{\textbf{v}}_3$ and the opening half-angle of this cone, we evaluate the 
properties of a vector $\textit{\textbf{v}}_2$ that extends from the Galactic center 
to the end point of $\textit{\textbf{v}}_3$, which is simply given by vector 
difference

\begin{equation}
\textit{\textbf{v}}_2=\textit{\textbf{v}}_3 -
\textit{\textbf{v}}_1=\left(\begin{array}{c} \mathcal{L} \cos \ell \cos b - R\\ \mathcal{L} \sin 
\ell \cos b\\ \mathcal{L} \sin b\end{array}\right)~.
\label{v2_noz}
\end{equation}

The length of this vector is given by

\begin{equation}\label{r_noz}
r=\sqrt{\mathcal{L}^2 - 2R\rho\cos \ell \cos b + R^2}~,
\end{equation}
and the half opening angle is given by
\begin{equation}\label{alpha_noz}
OA/2=\cos^{-1}(\mathcal{L}\sin b/r)~.
\end{equation}

If the gas parcel is traveling with a radial velocity $v_r$, the projected velocity along the line of sight is given by 
\begin{equation}
v_{GSR} \;=\; v_r \cos \beta.
\label{model_prediction}
\end{equation}     
Where $\beta$ is the angle subtended by the radial velocity vector with a vector from the sun towards the gas parcel and is defined as

\begin{equation}
\label{dot_product_unitv_noz}
\begin{split}
& \cos \beta \;=\; \hat{\textit{\textbf{v}}}_2\centerdot \hat{\textit{\textbf{v}}}_3=r^{-
1}\left(\begin{array}{c} \mathcal{L} \cos \ell \cos b - R\\ \mathcal{L} \sin \ell \cos b\\ \mathcal{L} \sin 
b\end{array}\right)\centerdot \left(\begin{array}{c}\cos \ell\cos b\\ \sin \ell\cos 
b \\ \sin b\end{array}\right)\\
&=(\mathcal{L} - R\cos\ell\cos b)/r~.
\end{split}
\end{equation}
This equation is algebraically identical to one defined earlier by \cite{Keeney2006}\footnote{There is a typographical error in this equation (Eq.~1) of \cite{Keeney2006} that was recognized and corrected by \cite{McClure-Griffiths2013}} 
but is much simpler in form. In the discussion that follows, we will compare the model predictions given by equation \ref{model_prediction} to the observed HVC kinematics, in the Local Standard of Rest velocities given by equation \ref{eqn_gsr_to_lsr}.

\subsection{Conical outflow from a circular zone}
The outflow from the interior of a circular zone of radius $r_{c}$ in the plane of the Galaxy can be characterized by a flow originating from a virtual point on the opposite side of the plane. This vertex point is located at a perpendicular distance $z\; = \; r_{c} \cot \alpha_{max}$ from the Galatic center. Where $\alpha_{max}$ is the half opening angle of the widest of the nested cones. To account for this displacement, we add a value of $-z$ to the (original zero) z-axis term in the expression of $\textit{\textbf{v}}_1$. We rewrite equation \ref{v2_noz} with an extra $z$ term,
\begin{equation}
\textit{\textbf{v}}_2=\left(\begin{array}{c} \mathcal{L} \cos \ell \cos b - R\\ \mathcal{L} \sin 
\ell \cos b\\ \mathcal{L} \sin b+z\end{array}\right)~,
\label{v2_noz_circ}
\end{equation}

and modify equations \ref{r_noz}, and \ref{alpha_noz} to read
\begin{equation}
r=\sqrt{\mathcal{L}^2 - 2R\rho\cos \ell \cos b + R^2 + 2 z \mathcal{L} \sin b + z^2}~,
\end{equation}
and 
\begin{equation}
OA/2=\cos^{-1}[(\mathcal{L}\sin b + z )/r]~.
\end{equation}

We rewrite equation \ref{dot_product_unitv_noz} for a conical outflow from a circular zone as 

\begin{equation}
 \cos \beta \;=\; \hat{\textit{\textbf{v}}}_2\centerdot \hat{\textit{\textbf{v}}}_3=(\mathcal{L} - R\cos\ell\cos b + z \sin b)/r~.
\end{equation}

\subsection{Wind Models}
We consider four wind models as described below. For each case we assume a singular isothermal sphere (SIS) density profile given as 
\begin{equation}
\rho (r) \; =\; \frac{ \sigma^{2} }{ 2\pi G r^2},
\end{equation}
where $\sigma$ is the velocity dispersion for the Milky Way halo. 

\textbf{Momentum driven:} We first consider a momentum driven wind model, where the outflow climbs ballistically out of the Galactic potential well after being given an initial impulse (e.g., by ram pressure from a hot wind). The equation of motion of such a wind is given by (see equation 13 of \citealt{Dijkstra2012}),
\begin{equation}
\frac{dv_{r}}{dt} \;=\; \frac{-G M(r)}{r^2} + A r^{-\alpha}  .
\end{equation}
We assume $\alpha$ = 2, which makes this equation equivalent to equation 24 of \cite{Murray2005}. The launch velocity is defined as $v_{L} \;=\; \sqrt{2Ar_{min}^{1-\alpha} /(\alpha -1)}$. For a bipolar conical outflow from the GC, we study two momentum driven models; the model (M1) is with $v_{L}\; \approx \; 1000$ {\kms}, and the second model (M2) is with $v_{L}\;\approx \; 1300$ {\kms}. Further, we study another model (S) that represents the geometry of a conical outflow from a circular zone of radius 200 pc around the GC. In this case we assume a launch velocity of  $v_{L}\;\approx \; 1000$ {\kms}. All the launch velocities are chosen to approximately bracket the observed velocity kinematics.

\textbf{Constant Energy:} We further define a model, that is launched with a constant energy explosion, model E for which we define the equation of motion as

\begin{equation}
\frac{4}{3} \pi \rho (r) v_{r}(r) ^{2} r^3 \;=\; \eta E_{0},
\end{equation}
where $E_{0}$ is the total energy of the explosion and $\eta$ is the efficiency parameter that controls the fraction of the total energy that drives the wind. We assume that  $E_{0} =6.7 \times 10^{55}$ erg, the total energy of the Fermi bubble (\citealt{Crocker2014} and $\eta$=1). 

\textbf{Constant Luminosity:} Lastly we study a model which drives a wind with constant luminosity, (model L), and the equation of motion is given as 
\begin{equation}
\frac{4}{3} \pi \rho (r) v_{r}(r) ^{3} r^2 \;=\; \eta_{0} L_{0},
\end{equation}
where $L_{0}$ is the total luminosity that is generated and $\eta_{0}$ is the efficiency parameter that controls the fraction of the total luminosity that drives the wind.  We assume $ L_{0}$ is the Eddington luminosity of the GC black hole, and $\eta_{0} $ = 2.5.

\begin{figure*}
\includegraphics[height=15cm, width= 18cm]{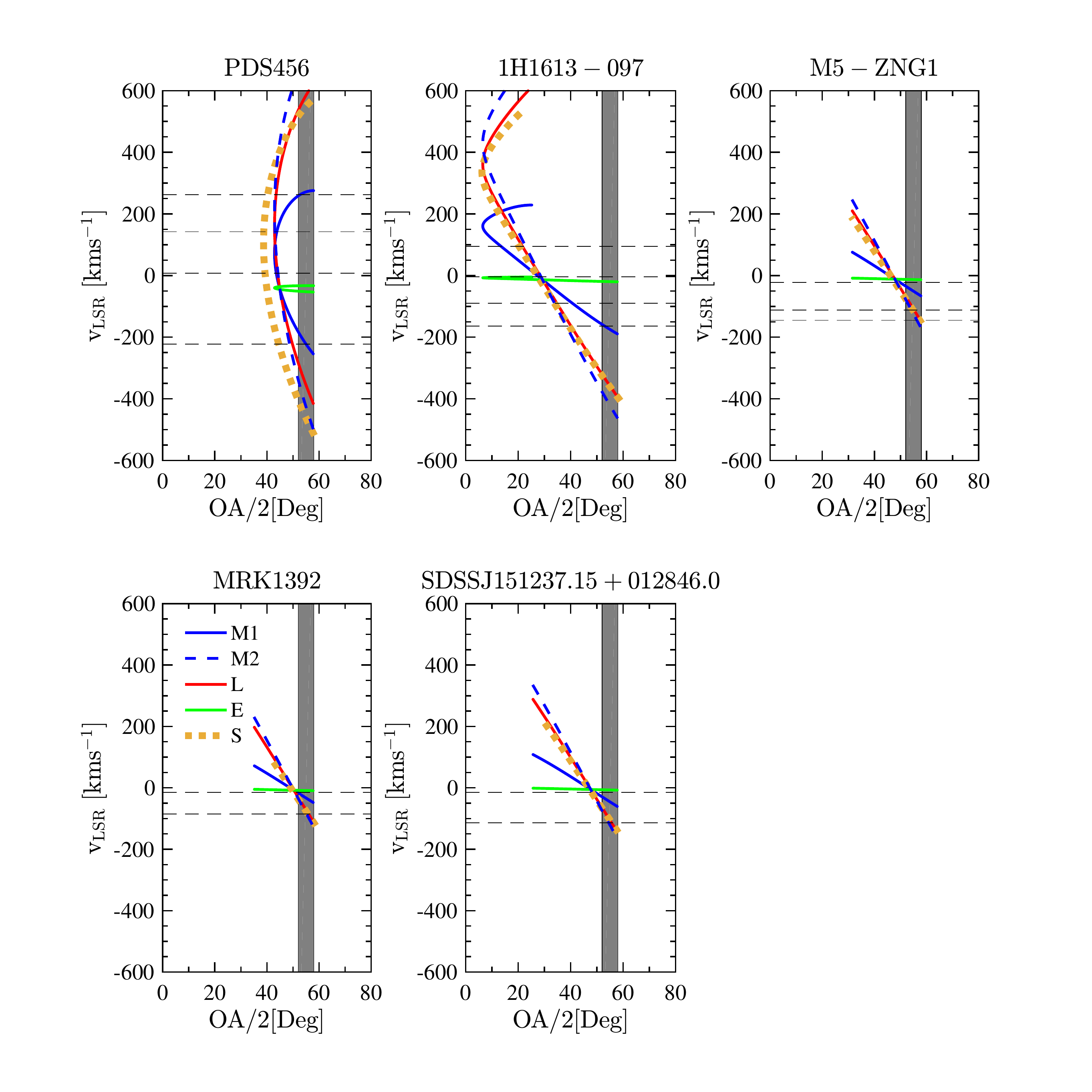}
  \caption{Kinematic models of the Galactic biconical nuclear outflow, some of which can explain the observed absorption-line centroids for the five lines of sight. The centroids of the observed absorption components in the five spectra are shown with horizontal dashed lines. The colored lines correspond to the predicted velocities as a function of distance from the sun, along that sightline. The colored lines at negative and positive $v_{LSR}$ correspond to the near side and the far side of the outflow bicone, respectively. $v_{LSR} \approx $ 0 corresponds to a distance directly above the GC. The gray band shows the opening angle range that matches the X-ray bicone. If a horizontal dashed line intersects a solid line (model prediction) to the left of the dashed band, the model predicts the kinematics of the observed absorption component along that line of sight.  Models M1 and M2 are momentum-driven models with launch velocities of $\approx$1000 and 1300 {\kms}, respectively. Model E is the constant energy and model L is the constant luminosity model, respectively. Model S is momentum driven  model with a launch velocity $\approx$ 1000 {\kms} from a circular zone of radius 200 pc around the GC.}
\label{fig:model_combined}
\end{figure*}

We compare these models to the observations along the five lines of sight inside the northern Fermi bubble. Figure \ref{fig:model_combined} shows the model $v_{LSR}$ velocities for the five lines of sight as a function of half opening angles of the bicone.  We use the {\CII} 1334 velocity centroids to compare the model to the observations (For redshifted PDS 456 component we use Si III 1206 as CII* contaminates the redshifted component). The horizontal dashed lines mark the centroids of the observed absorption along the five lines of sight. The gray band shows the half opening angle range of $\sim$ 55$^{\circ}$, that matches the X-ray bicone seen in the ROSAT data \citep{BlandHawthorn2003}. The intersection of the horizontal dashed lines with the colored model curves denote the model opening angles at which an observed absorption component exists. If a horizontal dashed line intersects a colored model curve (model prediction) to the \textit{left} of the dashed band then the observed absorption component kinematics along that line of sight can be explained by that model. In such a scenario, the cool outflowing gas is entrained inside the X-ray bicone and not along the edge of the X-ray bicone. If a horizontal dashed line intersects a colored model curve inside the vertical gray band, then the observed absorbing gas resides along the edge of the X-ray bicone. If the line intersections are to the right of the gray band,  or the model curves do not intersect the observed velocity ranges, then this model fails to reproduce the observed kinematics.

In all cases the constant energy model (E, green line) fails to represent the kinematics of the Fermi bubble. At high latitudes, this model predicts that any entrained outflowing gas would be observed at $v_{LSR}$ = 0 \kms, hence we rule out this constant energy model to explain the observed kinematics of the Fermi bubble. The momentum driven wind model M1 (solid blue line), satisfactory predicts the velocity ranges, where all the absorption components along PDS456 and 1H1613-097 are observed. However, M1 cannot reproduce the velocity ranges where the most blueshifted absorption components for M5-ZNG, SDSSJ151237.15+012846.0 and MRK1392 are seen. The Model M2 (dashed blue line) recovers the kinematics for all five lines of sight, and predicts extended blue and redshifted kinematics for the two low latitude lines of sight. In all the models discussed, we assume that the filling factor of entrained gas inside the bicone to be unity. In reality, the entrained cool gas clouds will be clumped together in small substructures. It is plausible that the reason that the  predicted extended blue and redshifted kinematics are not observed, is owing to the non-unity filling factor of entrained material inside the outflowing bicone.  The observational geometry makes these observations orthogonal to the traditional down-the-barrel observations of galactic winds, as we are probing the cool gas entrained in the nuclear outflow at different scale heights from the disk of the Milky Way, whereas in down-the-barrel observations the integrated effect of all outflowing gas is observed as blueshifted absorption/emission wings against the stellar continuum of the host galaxy.  The covering fraction computed for such observations gives the fraction of continuum source that is covered by the blueshifted outflowing gas (e.g. \citealt{Rupke2005}). These estimations are model dependent and can be estimated by solving the radiative transfer equations \cite{Hamann1997}. In \cite{Chisholm2016}, it was shown that for low ionized gas, the covering fraction of the outflowing gas in a local starburst galaxy is close to 100\%. It is comparable to the covering fraction estimated in section 4.1 where we find that the covering fraction of cool gas seen as blueshifted HVCs inside the FB is unity.  We are probing a radial distance of up $\approx$ 7 kpc from the disk of the MW.  Hence we are observing similar gas covering fraction as seen in the case of down-the-barrel spectroscopy of starburst galaxies in the local Universe. However, both down-the-barrel and our orthogonal observations cannot yet constrain the detailed kinematic structure of the outflowing gas (e.g.  filling factor of entrained material inside the outflowing bicone). It is also probable that there were two separate events of momentum injection which happened $\approx$ 6 and 4 Myrs ago. The first event represented by M2 can explain the kinematics of M5-ZNG1 and MRK1392 and the second event represented by M1 can explain the kinematics of PDS456 and 1H1613-097, respectively.

Both the constant luminosity model (L, red line), and the model of conical outflow from a circular zone (S, dotted orange line), can recover the observed kinematics of all the five lines of sight, and again predicts extended blue and redshifted kinematics for the two low latitude lines of sight. These predicted extended kinematics might not be observed owing to non-unity filling factor of entrained cool gas in the outflow. The constant luminosity model (L) argues for a super Eddington luminosity AGN being active in the GC for $\approx$ 5-6 Myrs, and the model of conical outflow from a circular zone (S, dotted orange line) predicts an event of momentum injection which happened $\approx$ 6$-$7 Myrs ago.

For each model we have a radial profile, and we can compute the time taken for outflowing gas to be launched from the Galactic Center to any arbitrary radial distance R. This time is computed as 
\begin{equation}
 T = \int_{0}^{R} 1/v_{r} \, dr. 
 \end{equation}
Again assuming that the opening angle of the outflowing gas is well represented by the half opening angle of the X-ray bicone, we compute the mean radial distance from the GC to each line of sight. These are tabulated in Table \ref{table:Vpfit_measurements}. 

\begin{figure*}
\centering
\includegraphics[height=7.5cm,width=8.9cm]{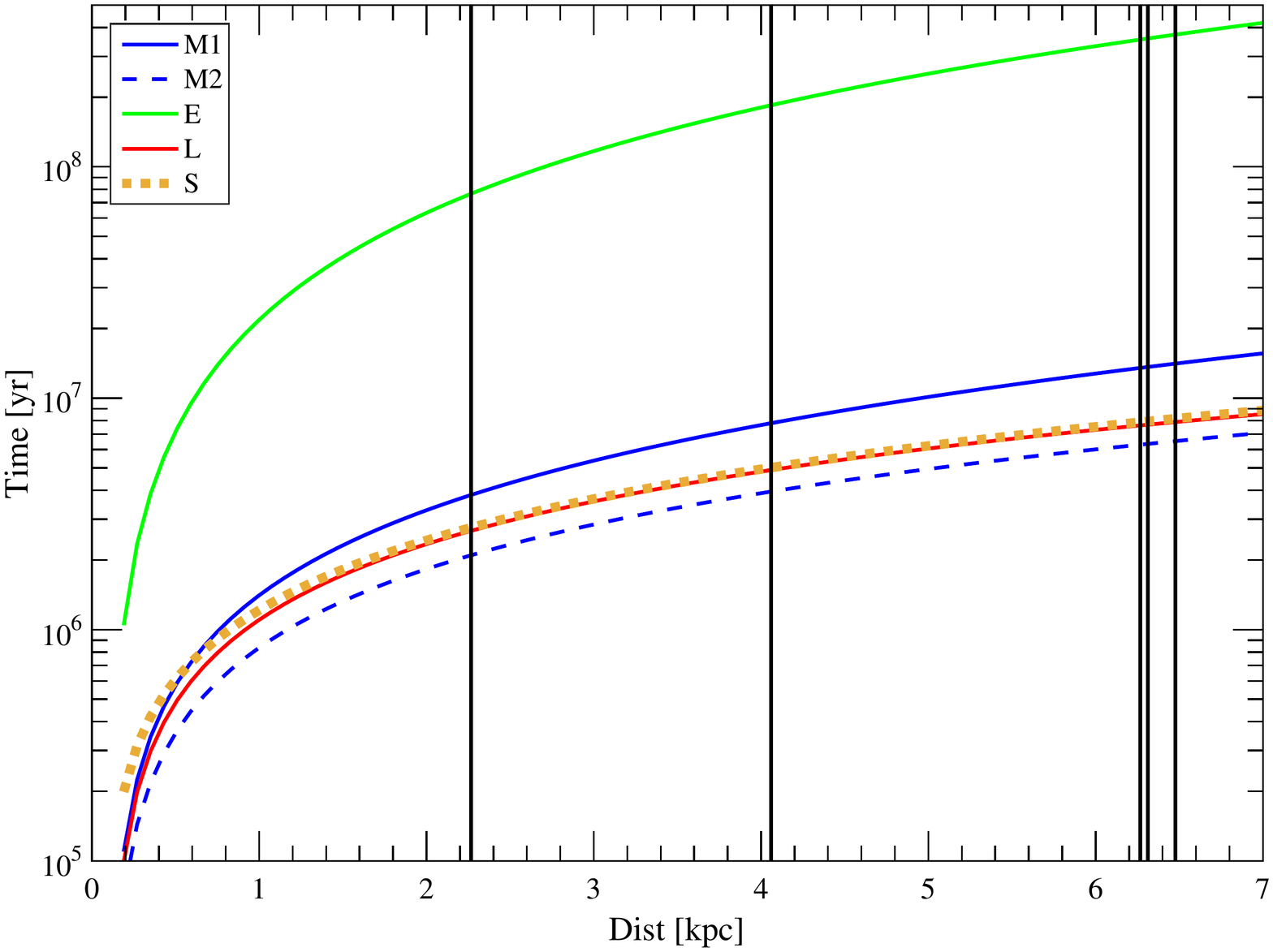}
\includegraphics[height=7.5cm,width=8.9cm]{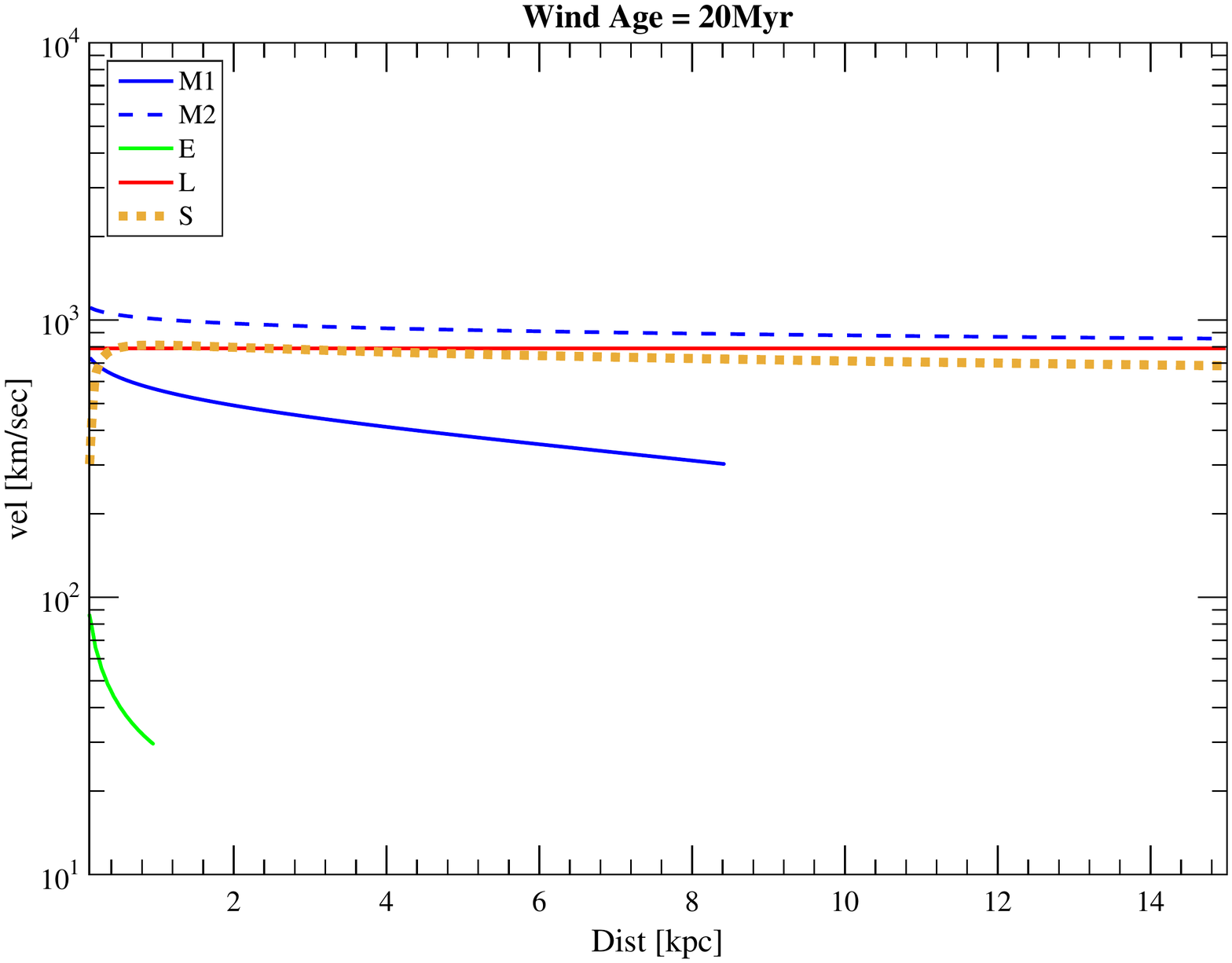}
  \caption{Left Panel: The time taken for the five outflow models as a function of radial distance along the outflow cone. The vertical lines show the mean radial distance of the outflowing gas parcel from the GC. The momentum drive wind models (M1, M2), outflow from a circular zone around GC (S), and the constant luminosity model (L) will  need to drive the outflow for $\approx$ 6 to 9 Myrs to reach the absorption seen in MRK1392. The constant energy model (E) will  need to drive the outflow for $\approx$ 400 Myrs to reach the absorption seen in MRK1392. Right Panel: The radial profile of the four outflow models after driving outflows for 20 Myrs.}
\label{fig:model_time}
\end{figure*}

The left panel of Figure \ref{fig:model_time} shows the time taken by the outflowing gas to reach any radial distance from the GC. The mean radial distance to the five lines of sight are shown as vertical dashed lines. The momentum driven wind models (M1, M2), the conical outflow model from a circular zone (S), and the constant luminosity model (L) will need to drive the outflow for $\approx$ 6 to 9 Myrs to reach the absorption seen in MRK1392. 

Figure \ref{fig:model_time}, right panel shows the radial distance traveled by the outflowing gas, after 20 Myrs. The outflowing gas launched by the constant energy model E has not travelled sufficient radial distance to reach PDS456 in 20 Myrs. All the other wind models have driven the outflowing gas enough to reach all the five lines of sight studied here. The radial profile of the constant energy model E falls off sharply at 1.5 kpc. Even if the outflows driven by this model reaches the four lines of sight, their projected velocities will be close to zero (see Figure \ref{fig:model_combined}). Hence, we can rule out model E to be driving the outflows observed in the Fermi Bubble.   

Figure \ref{fig:radial_absorption_profile}, show the radial velocity profiles (left panel) and absorption profiles (right panel) inside the northern FB as a function of mean de-projected radial distance from the GC. The symbols are the same as in Figure \ref{fig:radial_profile}. These profiles show the de-projected radial kinematic and absorption profile of gas inside the northern FB from 2.3 kpc from the GC to 6.5 kpc.

These models and the kinematic observations do not allow us to distinguish between an AGN driven or star-formation driven origin theory of the Fermi Bubble. But they provide us with an independent measure of the age of the Fermi Bubble. We find that any energetic event that created the Fermi Bubble must have been short ($\sim$ 6 $-$9 Myrs). Hence we can rule out a sustained $> 10^{8}$ yr star-formation being the origin of the Fermi Bubble. Recent works have shown evidence of both AGN and starburst driven winds from the ionization structure around local galaxies \citep{Sharp2010}.  Recently, \cite{Miller2016} used X-ray data and independently estimated the age of the Fermi Bubbles to be $\approx$ 4.3 Myrs. With completely different methods, we are estimating a very similar Fermi Bubble age as in  \cite{Miller2016}.

In this section, we restrict our analysis to compare these models with observations along lines of sight that are only passing through the northern Fermi bubble. Our models are too simplistic to explain the complex kinematics that might be associated with the boundary regions of the Fermi bubbles. The complex gas kinematics seen along these sightlines would be dominated by shocks from the nuclear outflow terminating in those regions and local instabilities \citep{Thompson2016,Bordoloi2016b}. Any modeling of such complex physical processes would require more  sophisticated modeling and is beyond the scope of this work.


\begin{figure*}
\includegraphics[height=6.cm,width=8.75cm]{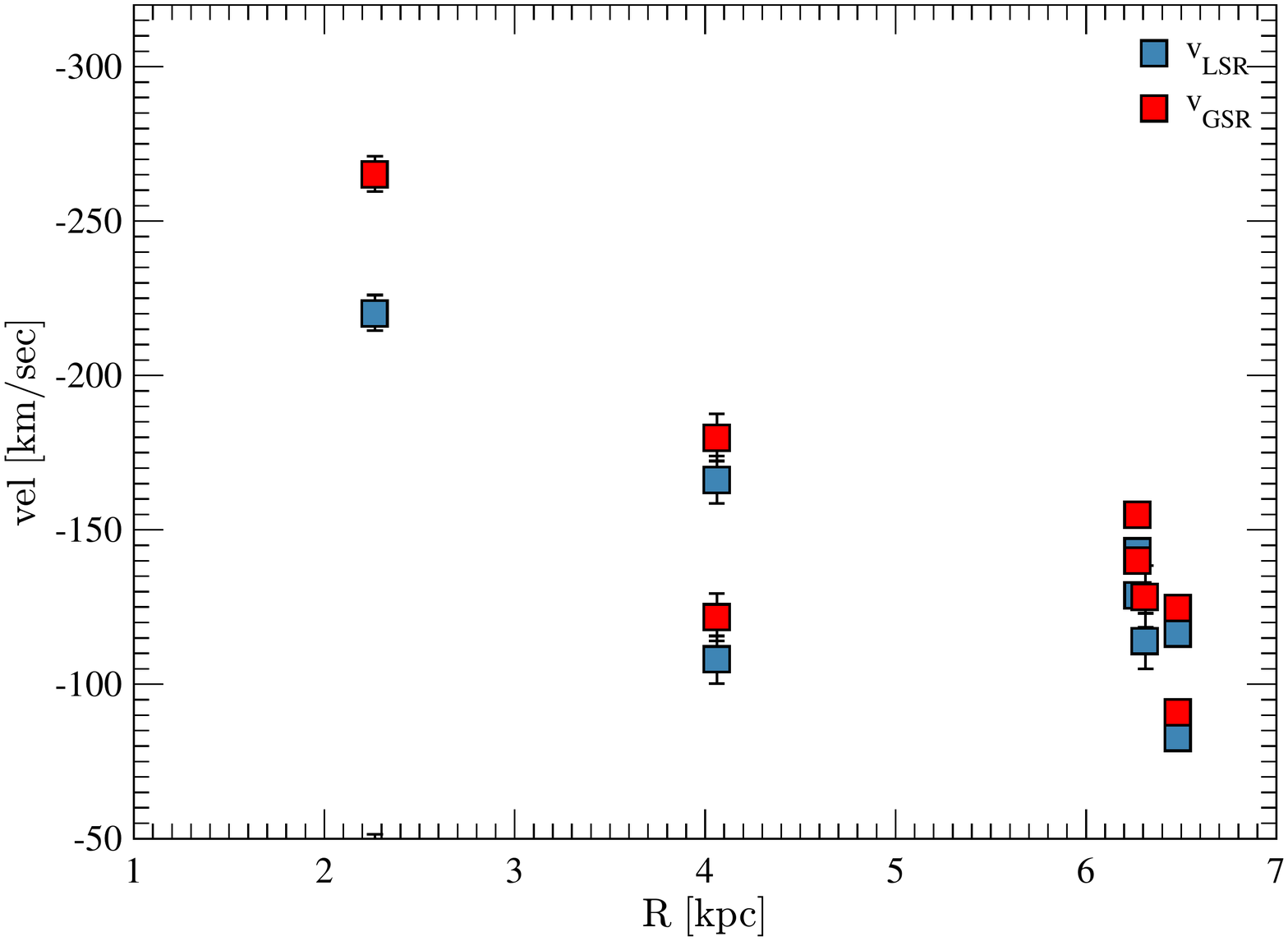}
\includegraphics[height=6.cm,width=8.75cm]{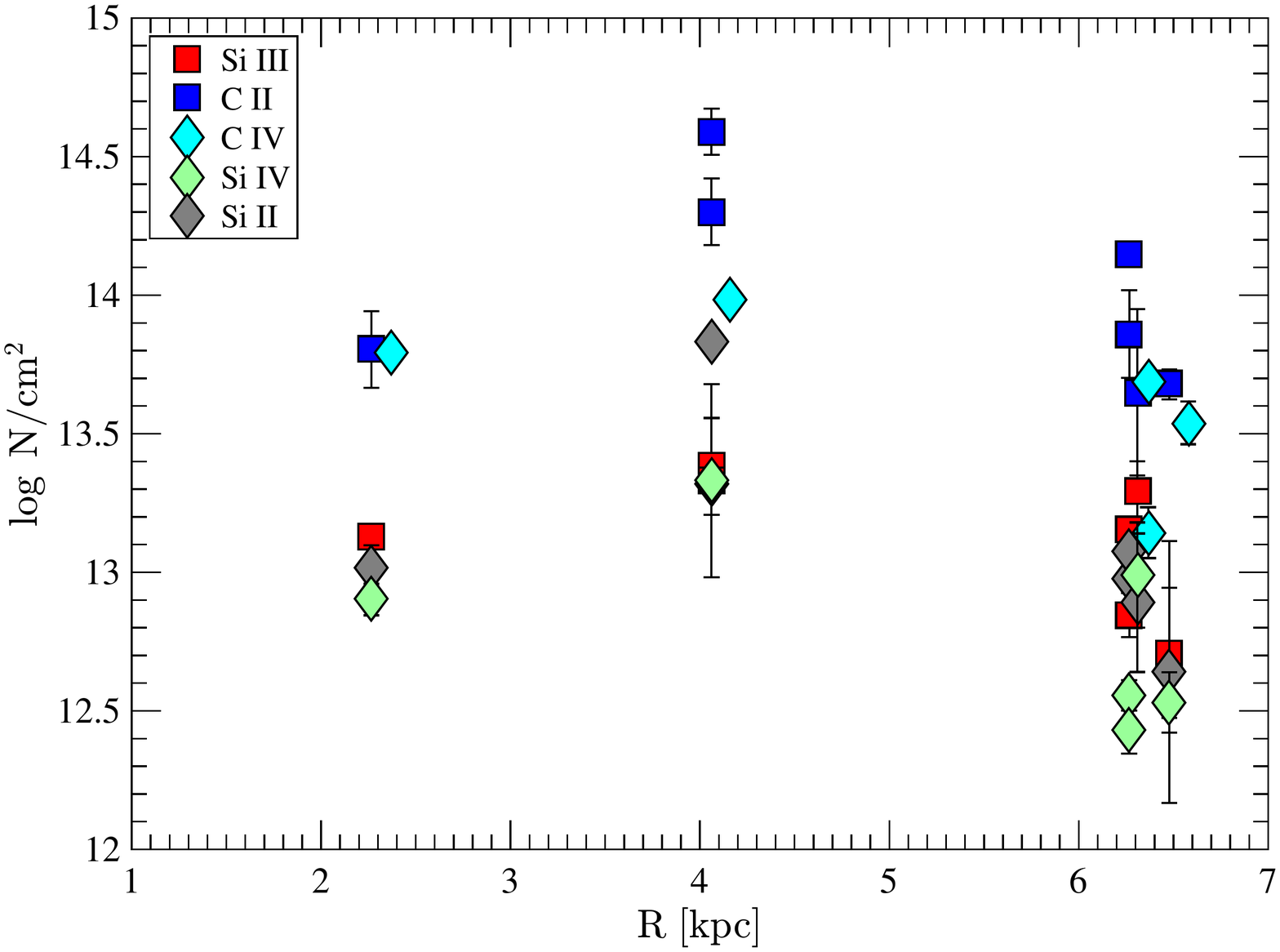}
  \caption{ Left Panel: Radial velocity profile of the blueshifted {\CII} high velocity absorption inside the Fermi Bubble as a function of mean radial distance from the GC.   \textit{Right Panel:}  Radial absorption profile of the blueshifted high velocity absorption inside the Fermi Bubble as a function of mean radial distance from the GC. In both panels the symbols are same as Figure \ref{fig:radial_profile}. }
\label{fig:radial_absorption_profile}
\end{figure*}	

\section{Entrained Mass in the Fermi Bubble} 

The amount of gas ejecta carried out by the nuclear outflow is of great importance as such mass flows are believed to be regulating the evolution of galaxies. Given that we can constrain the half opening angle of the nuclear outflow from the X-ray bicone to be $\alpha \; \approx$ 55$^{\circ}$, we can estimate the cold gas, mass outflow rate for such a scenario. From Section 5, we know the average distance of the gas clouds from the GC, and the de-projected radial velocities of the outflowing gas. Following \cite{Bouche2011}, we can express the mass outflow rate at a distance $b$ kpc from the GC as 

\begin{equation}
\dot{M}_{out}(b) \;=\; \frac{\pi}{2} \alpha \; N_{H} \; b\;  v_{r}.
\label{Mdot_eqn}
\end{equation}

Here $N_{H}$ is the total H column density. Inserting numerical values, equation \ref{Mdot_eqn} can be rewritten as 
\begin{equation}
\rm{\dot{M}_{out}(b) \;  =\; 0.41 \; M_{\odot}\; \rm{yr^{-1}}\;   \frac{\mu}{1.5} \;\frac{\alpha}{30^{\circ}} \; \frac{N_{H}(b)}{10^{19}\; \rm{cm^{-2}}} \\
 \; \frac{b}{25\; \rm{kpc}}\;  \frac{v_{r}}{200\; \rm{kms^{-1}}},}
\label{Mdot_eqn2}
\end{equation}

where $\mu$ is the mean atomic weight. We stress that these mass flow rate estimates are highly uncertain and model dependent, and should only be taken as rough back of the envelope calculation, prone to systematic uncertainties. Keeping this caveat in mind, we report the minimum mass flow rates below.

Following Section 4; for the most blueshifted absorption component of 1H1613-097, we assume the most conservative value of $v_{r}$ from the momentum driven wind scenario (M1), and get $v_{r}$ = 410 {\kms} at a mean radial distance of $b$ = 4.06 kpc. To estimate the minimum $N_{H}$ column density for all the high velocity clouds along this line of sight, we sum the HVC Si column densities (i.e. $N_{\rm{Si}}  \; =\; N_{\rm{SiII}} \;+\; N_{\rm{SiIII}}\;+\; N_{\rm{SiIV}}$). By summing over Si stages II + III + IV, we effectively are doing an ionization correction, and assuming that all the Si atoms exist in these phases. The total $N_{H}$ column density is given as  $\log (N_{H}) = \log( N_{Si}) + 0.54 + 4.49$, where we assume that  [Si/H] $\approx$ [O/H] $\approx$ $-$0.54 (without any correction to account for depletion onto dust grains), and (Si/H)$_{\odot}$ = $-$4.49 \citep{Asplund09}. This yields a total hydrogen column density of $N_{H}\; \gtrsim \; 1.1 \times10^{19}$ \pcm. Inserting these values into equation \ref{Mdot_eqn2} yields a minimum mass outflow rate of $\dot{M}_{out} \gtrsim $ 0.22 $\rm{ M_{\odot}\; yr^{-1}}$.  We also estimate the minimum momentum flux carried out by the outflowing gas as $\dot{P}_{out} \gtrsim \dot{M}_{out} v_{out}$ dynes. For the 1H1613-097 sightline the minimum momentum flux carried out by the outflowing gas is $ \gtrsim 7.2 \times 10^{32}$ dynes.

Similarly, we estimate the minimum mass outflow rate along the PDS456 direction to be $\dot{M}_{out} \gtrsim$  0.2 $\rm{ M_{\odot}\; yr^{-1}}$. Here also we assume the most conservative value of $v_{r}$ = 479 {\kms} from the momentum driven wind scenario (M1), at a mean radial distance of $b$=2.27 kpc.  We add up all the HVC Si column densities as before, and estimate a total hydrogen column density of  $N_{H}\; \gtrsim\; 1.3 \times10^{19}$ \pcm. For the PDS456 sightline the minimum momentum flux carried out by the outflowing gas is $\gtrsim 7.6 \times 10^{32}$ dynes. Table \ref{mass_outflow_rate} shows these mass outflow rate estimates. 

If we assume that the Fermi Bubble has existed for at least the amount of time it takes for outflow to reach MRK1392, the highest latitude of our inside-the-Bubble sightlines, we get the most conservative age from the M2 model of $\approx$ 6 Myrs. Now if we assume that there was an outflow for that duration of time with an average mass outflow rate $\gtrsim$  0.2 $\rm{ M_{\odot}\; yr^{-1}}$; then assuming mirror symmetry, the total minimum mass of  cool gas entrained in both  the Fermi Bubbles is $\gtrsim \rm{2 \times10^{6} M_{\odot}}$.

As we discuss in the previous section, we cannot distinguish between AGN and SF from pure kinematics alone. However, these kinematics allow us to constrain the outflow velocity and mass flow rates of the FB. Our kinematic age estimate shows that the event that created the FB must be short ($<$ 6$-$9 Myr). Knowing the total entrained mass and outflow velocity of the wind allows us to estimate the total kinetic energy associated with the FB to be $\rm{E_{kin} \sim (1/2)\times Mass\times Vel^{2}}$ $\rm{\sim (2 \times10^6)\times(1300^2) \sim 6\times10^{55}}$ ergs. Though it is very rough back of the envelope calculation, this energy budget indeed argues for AGN activity powering the bubbles.

\section{Summary}
In this paper we have studied the kinematics and properties of the entrained gas inside the northern Fermi Bubble as traced by the UV absorption lines of O I,  Al II, C II, C IV, Si II, Si III, Si IV and other species. This analysis is based on a sample of 46 extragalactic sightlines observed with HST/COS and one sightline observed with HST/STIS. This is the first work that fully characterizes the velocity profile and spatial extent of the entrained absorbing material driven by the nuclear outflow from the Galactic Center. The main findings of this work are as follows:

\begin{itemize}
\item All five lines of sight passing through the Fermi Bubble exhibit blueshifted absorption whereas 9 out of the 42 lines of sight outside the Fermi Bubble exhibit blueshifted absorption. Inside the Fermi Bubble, only PDS456 at low galactic latitudes exhibits both blueshifted and redshifted high velocity absorption components, which can be understood as tracing the front and back side of the outflow.  The incidence of any (blueshifted or redshifted) high velocity absorption inside the Fermi Bubble is 92 $\pm$ 8\% (5/5). The incidence of blueshifted high velocity absorption outside the Fermi Bubble is 22 $\pm$ 6\% (9/42) and any high velocity absorption is 31 $\pm$ 7\% (13/42). For all cases,   adjusted Chi-squared tests show that we can rule out the null hypothesis that the distribution of blueshifted high velocity absorbers inside and outside the northern Fermi bubble are the same at more than 99.8\% confidence level.

\item We characterize the observed velocity profile of the outflowing gas as a function of Galactic latitude and the de-projected radial distance from the Galactic Center. We observe a monotonically decreasing blueshifted outflow velocity with increasing Galactic latitude and distance. The observed blueshifted velocities change from $v_{GSR} =$ $-$265 {\kms} at a radial distance of 2.3 kpc to $v_{GSR} =$ $-$91 {\kms} at a radial distance of 6.5 kpc. This spatial constraint matches the Fermi Bubble observed in gamma-ray emission.

\item By combining HST/COS observations of O I 1302 with a Green Bank Telescope detection of H I 21 cm emission, we estimate the metallicity of the blueshifted HVC along the 1H1613-097 sightline, finding  [O/H]  $\gtrsim -0.54 \pm 0.15$. This implies that the derived metallicity is $\gtrsim$ 30\% solar. 

\item We develop simple kinematic bipolar outflow models to explain the observed kinematics of the high velocity clouds inside the Fermi Bubble. We rule out a constant energy explosion model as the origin of the nuclear outflow. We find that two momentum injection events at $\sim$ 4 Myr and $\sim$6 Myr ago with launch velocities $\approx$ 1000 and 1300 {\kms} can satisfactorily explain the kinematics of all the four lines of sight inside the Fermi Bubble.  Alternatively, a constant luminosity AGN active at the GC for $\approx$ 5$-$6 Myrs can also explain the observed kinematics of the Fermi Bubble. 

\item These kinematic models constrain the age and spatial extent of the UV-absorbing gas within the Fermi Bubble to be $\approx$ 6-9 Myrs, and $\approx$ 6.5 kpc, respectively. Therefore the UV-absorbing gas seems to be confined to the same spatial regions as the gamma-ray emitting gas, even though it traces a very different temperature plasma.

\item Using the observed metal column densities and velocities, and a simple kinematic model, we estimate the minimum mass outflow rate from the nuclear outflow to be $\gtrsim$ 0.2 $\rm{ M_{\odot}\; yr^{-1}}$. The total minimum mass of cool gas entrained in both the Fermi Bubble is $\gtrsim \rm{2 \times10^{6} M_{\odot}}$.
\end{itemize}

\section{Acknowledgement}
Support for this work was provided by NASA through Hubble Fellowship grant \#51354 awarded by the Space Telescope Science Institute, which is operated by the Association of Universities for Research in Astronomy, Inc., for NASA, under contract NAS 5-26555. Support for program 13448 was provided by NASA through grants from the Space Telescope Science Institute, which is operated by the Association of Universities for Research in Astronomy, Inc., under NASA contract NAS 5-26555. The Green Bank Observatory is a facility of the National Science Foundation operated under cooperative agreement by Associated Universities, Inc and the data for this project were obtained through the Green Bank Telescope under the NRAO program ID GBT/14B-299. TSK acknowledges funding support from the European Research Council Starting Grant `Cosmology with the IGM' through grant GA-257670.

\clearpage
\LongTables			
\begin{deluxetable*}{ccccccc}
\tablecolumns{7}
\tablewidth{0pt}
\tablecaption{List of Targets in and around the Northern Fermi bubble }
\tablehead{
			\colhead{Object \#}                          &
			\colhead{Name}								 &
			\colhead{ $l$ [Deg]} 						 &
			\colhead{ $b$ [Deg]}						 &
			\colhead{ $\rm{z_{QSO}}$}					 &
			\colhead{ Location wrt FB \tablenotemark{a}} &
			\colhead{ GBT spectrum \tablenotemark{b}}    
}
\startdata
 1   &   PDS456                       &      10.4      &      11.2         &                       0.1840   &  Inside     &     Yes \\  
 2   &   QSO1500-4140                 &      327.7     &      14.6         &                       0.3350   &  Outside    &     No \\  
 3   &   1H1613-097                   &      3.5       &      28.5         &                       0.0650   &  Inside     &     Yes \\  
 4   &   M5-ZNG1                      &      3.9       &      47.7         &    Halo Star\tablenotemark{c}   &  Inside     &     Yes \\  
 5   &   PG1709+142                   &      34.9      &      28.5         &    Halo Star\tablenotemark{d}   &  Outside    &     No \\  
 6   &   MRK877                       &      32.9      &      41.1         &                       0.1124   &  Outside    &     No \\  
 7   &   LBQS1435-0134                &      348.7     &      51.4         &                       1.3077   &  Interface   &     No \\  
 8   &   RX\_J1605.3+1448             &      27.8      &      43.4         &                       0.3721   &  Outside    &     No \\  
 9   &   PG1522+101                   &      14.9      &      50.1         &                       1.3210   &  Interface    &     No \\  
10   &   PG1553+113                   &      21.9      &      44.0         &                       0.4700   &  Interface   &     No \\  
11   &   SDSSJ154553.50+093620.0      &      18.3      &      45.4         &                       0.6650   &  Interface   &     No \\  
12   &   PG1435-067                   &      344.0     &      47.2         &                       0.1260   &  Interface   &     No \\  
13   &   SDSSJ151237.15+012846.0      &      1.8       &      47.5         &                       0.2650   &  Inside     &     No \\  
14   &   3C323.1                      &      33.9      &      49.5         &                       0.2653   &  Outside    &     No \\  
15   &   MRK1392                      &      2.8       &      50.3         &                       0.0363   &  Inside     &     Yes \\  
16   &   SDSSJ151507.40+065708.0      &      9.0       &      50.4         &                       0.2680   &  Interface   &     No \\  
17   &   SDSSJ150952.20+111047.0      &      13.6      &      53.8         &                       0.2849   &  Outside    &     No \\  
18   &   RBS1454                      &      5.6       &      52.9         &                       0.2860   &  Interface   &     No \\  
19   &   SDSSJ150928.30+070235.0      &      7.8       &      51.6         &                       0.4188   &  Interface   &     No \\  
20   &   MRK841                       &      11.2      &      54.6         &                       0.0364   &  Outside    &     No \\  
21   &   SDSSJ142614.79+004159.4      &      347.6     &      55.1         &                       0.8950   &  Outside    &     No \\  
22   &   RX\_J1429.6+0321             &      351.8     &      56.6         &                       0.2530   &  Outside    &     No \\  
23   &   SDSSJ145450.10+111434.0      &      10.2      &      57.0         &                       0.4681   &  Outside    &     No \\  
24   &   SDSSJ140655.66+015712.8      &      341.8     &      59.0         &                       0.4270   &  Outside    &     No \\  
25   &   HE1340-0038                  &      328.8     &      59.4         &                       0.3260   &  Outside    &     No \\  
26   &   SDSSJ141949.40+060654.0      &      351.9     &      60.3         &                       1.6380   &  Outside    &     No \\  
27   &   SDSSJ135726.27+043541.4      &      340.8     &      62.5         &                       1.2340   &  Outside    &     No \\  
28   &   RX\_J1342.1+0505             &      333.9     &      64.9         &                       0.2660   &  Outside    &     No \\  
29   &   RX\_J1426.2+1955             &      19.6      &      67.2         &                       0.2100   &  Outside    &     No \\  
30   &   SDSSJ141542.90+163414.0      &      8.8       &      67.8         &                       0.7430   &  Outside    &     No \\  
31   &   PG1424+240                   &      29.5      &      68.2         &                       0.5000   &  Outside    &     No \\  
32   &   KUV14189+2552                &      33.8      &      69.9         &                       1.0530   &  Outside    &     No \\  
33   &   NGC5548                      &      32.0      &      70.5         &                       0.0170   &  Outside    &     No \\  
34   &   SDSSJ141038.40+230447.0      &      24.6      &      71.6         &                       0.7960   &  Outside    &     No \\  
35   &   SDSSJ135712.60+170444.0      &      2.9       &      71.8         &                       0.1500   &  Outside    &     No \\  
36   &   PG1352+183                   &      4.4       &      72.9         &                       0.1520   &  Outside    &     No \\  
37   &   PKS1354+19                   &      9.0       &      73.0         &                       0.7200   &  Outside    &     No \\  
38   &   RX\_J1356.4+2515             &      29.3      &      75.3         &                       0.1650   &  Outside    &     No \\  
39   &   RX\_J1342.7+1844             &      0.2       &      75.5         &                       0.3820   &  Outside    &     No \\  
40   &   SDSSJ135424.90+243006.3      &      25.9      &      75.6         &                       1.8920   &  Outside    &     No \\  
41   &   SDSSJ131545.20+152556.0      &      329.9     &      77.0         &                       0.4490   &  Outside    &     No \\  
42   &   SDSSJ134822.30+245650.0      &      26.4      &      77.0         &                       0.2930   &  Outside    &     No \\  
43   &   PG1341+258                   &      28.7      &      78.2         &                       0.0870   &  Outside    &     No \\  
44   &   SDSSJ131802.10+262830.0      &      28.2      &      84.0         &                       1.2350   &  Outside    &     No \\  
45   &   HS1302+2510                  &      357.4     &      86.3         &                       0.6020   &  Outside    &     No \\  
46   &   SDSSJ125846.70+242739.0      &      335.1     &      86.9         &                       0.3710   &  Outside    &     No \\  
47   &   RX\_J1303.7+2633             &      21.8      &      87.2         &                       0.4370   &  Outside    &     No \\
   \vspace{-0.2cm}
\enddata
\tablenotetext{a}{Whether the line of sight is inside the Fermi Bubble.}
\tablenotetext{b}{ Whether deep GBT HI 21 cm spectrum was obtained.}
\tablenotetext{c}{Distance from Sun = 7.5 kpc.}
\tablenotetext{d}{Distance from Sun = 21 kpc.}
\label{QSO_list}
\end{deluxetable*}


\clearpage 
\LongTables
\begin{deluxetable*}{cccc}
\tablecolumns{4}
\tablewidth{0pt}
\tablecaption{Voigt profile fit parameters for absorbers within the Fermi Bubble}
\tablehead{
\colhead{ Ion}&
\colhead{ $v_{LSR}$ [{\kms}] \tablenotemark{a}} &
\colhead{ $b$ [{\kms}]}&
\colhead{$\log N \; [\rm{cm^{-2}}]$  \tablenotemark{b}}
}
\startdata

  &                         &       PDS456         &         $\langle R \rangle$\footnote{Mean radial distance from the GC}= 2.27 kpc                     \\
\hline
{\SiIII}& -197\footnote{Ly-$\beta$ contamination at $z=0.175539$}  $\pm$ 2 & 39.3 $\pm$ 2.3 & 13.13 $\pm$ 0.02 \\ 
{\SiIII} & -4 $\pm$ 2 & 69.2 $\pm$ 2.8 & $>$14.04 \\ 
{\SiIII} & 145 $\pm$ 2 & 28.9 $\pm$ 2.8 & 13.06 $\pm$ 0.04 \\ 
{\SiIII} & 259 $\pm$ 2 & 18.2 $\pm$ 2.2 & 12.85 $\pm$ 0.04 \\ 
{\SiIV} & -231 $\pm$ 2 & 19.0 $\pm$ 3.4 & 12.90 $\pm$ 0.06 \\ 
{\SiIV} & 1 $\pm$ 1 & 45.0 $\pm$ 1.4 & $>$14.05 \\ 
{\SiIV} & 146 $\pm$ 32 & 107.7 $\pm$ 44.6 & 13.03 $\pm$ 0.16 \\ 
{\SiII} & -223 $\pm$ 2 & 13.2 $\pm$ 3.9 & 13.02 $\pm$ 0.08 \\ 
{\SiII} & 6 $\pm$ 1 & 45.4 $\pm$ 0.9 & $>$14.76 \\ 
{\SiII} & 122 $\pm$ 2 & 24.9 $\pm$ 2.0 & 13.40 $\pm$ 0.03 \\ 
{\SiII} & 264 $\pm$ 2 & 35.5 $\pm$ 2.2 & 13.37 $\pm$ 0.02 \\ 
{\CII} & -220 $\pm$ 6 & 19.9 $\pm$ 8.0 & 13.80 $\pm$ 0.14 \\ 
{\CII} & -2 $\pm$ 4 & 51.3 $\pm$ 9.9 & $>$15.44  \\ 
{\CII} & 123 $\pm$ 5 & 20.8 $\pm$ 6.7 & $>$14.14  \\ 
{\CIV} & -233 $\pm$ 2 & 31.9 $\pm$ 2.1 & 13.79 $\pm$ 0.03 \\ 
{\CIV} & 0 $\pm$ 1 & 51.0 $\pm$ 1.1 & $>$14.71  \\ 
{\CIV} & 147 $\pm$ 6 & 58.4 $\pm$ 9.4 & 13.58 $\pm$ 0.06 \\ 
{\AlII} & 8 $\pm$ 1 & 46.4 $\pm$ 2.0 & 13.59 $\pm$ 0.03 \\ 
{\AlII} & 124 $\pm$ 2 & 6.7 $\pm$ 5.2 & 12.12 $\pm$ 0.11 \\ 
{\AlII} & 263 $\pm$ 3 & 0.8 $\pm$ 3.0 & 13.53 $\pm$ 0.63 \\ 

\hline
        &                       &       1H1613-097         &        $\langle R \rangle$= 4.06 kpc                   \\
\hline        
{HI}\footnote{ {\HI} measurements are from the 21 cm observations.}    &  -172.2 $\pm$ 0.1 &         12.8 $\pm$ 0.8             &  18.23 $\pm$ 0.03 \\
{\SiIII} & -4 $\pm$ 8 & 38.0 $\pm$ 6.5 & $>$13.94  \\ 
{\SiIII} & -164 $\pm$ 12 & 35.5 $\pm$ 8.3 & $>$13.38  \\ 
{\SiIII} & -90 $\pm$ 13 & 38.7 $\pm$ 32.2 & $>$13.33  \\ 
{\SiIV} & -8 $\pm$ 1 & 29.5 $\pm$ 1.5 & $>$13.73 \\ 
{\SiIV} & -119 $\pm$ 5 & 59.3 \footnote{$b$ parameter fixed with {\CIV} value for better fit}  & 13.33 $\pm$ 0.04 \\ 
{\SiII} & -176 $\pm$ 1 & 22.1 $\pm$ 1.5 & 13.83 $\pm$ 0.03 \\ 
{\SiII} & -110 $\pm$ 2 & 22.8 $\pm$ 3.1 & 13.32 $\pm$ 0.04 \\ 
{\SiII} & -10 $\pm$ 1 & 37.4 $\pm$ 0.9 & $>$14.53  \\ 
{\CII} & -162 $\pm$ 5 & 31.3 $\pm$ 5.7 & $>$14.59  \\ 
{\CII} & -100 $\pm$ 4 & 18.9 $\pm$ 6.6 & $>$14.30  \\ 
{\CII} & -6 $\pm$ 3 & 35.9 $\pm$ 8.2 & $>$15.10  \\ 
{\CII} & 83 $\pm$ 4 & 14.9 $\pm$ 5.2 & 13.88 $\pm$ 0.11 \\ 
{\CIV} & -140 $\pm$ 3 & 59.3 $\pm$ 5.1 & 13.98 $\pm$ 0.03 \\ 
{\CIV} & 7 $\pm$ 1 & 29.7 $\pm$ 1.6 & $>$14.19  \\ 
{\OI} & 13 $\pm$ 1 & 31.7 $\pm$ 2.6 & $>$15.00  \\ 
{\OI} & -32 $\pm$ 15 & 67.5 $\pm$ 8.9 & 14.63 $\pm$ 0.13 \\ 
{\OI} & -163 $\pm$ 1 & 17.6 $\pm$ 1.7 & 14.28 $\pm$ 0.03 \\ 
{\AlII} & -152 $\pm$ 4 & 54.1 $\pm$ 5.5 & 12.85 $\pm$ 0.04 \\ 
{\AlII} & 0 $\pm$ 1 & 33.0 $\pm$ 1.6 & $>$13.41 \\ 
{\FeII} & -182 $\pm$ 3 & 7.6 $\pm$ 6.2 & 13.61 $\pm$ 0.13 \\ 
{\FeII} & -1 $\pm$ 2 & 28.2 $\pm$ 2.2 & $>$14.64 \\ 
{\NI} & -174 $\pm$ 3 & 32.1 $\pm$ 4.2 & 13.92 $\pm$ 0.05 \\ 
{\NI} & 22 $\pm$ 3 & 27.3 $\pm$ 4.2 & 13.93 $\pm$ 0.06 \\  \hline
        &                       &       M5-ZNG1         &              $\langle R \rangle$= 6.26 kpc             \\
\hline
{\SiIII} & -142 $\pm$ 1 & 7.1 $\pm$ 1.2 & 12.85 $\pm$ 0.08 \\ 
{\SiIII} & -118 $\pm$ 1 & 16.5 $\pm$ 1.3 & 13.15 $\pm$ 0.04 \\ 
{\SiIII} & -24 $\pm$ 1 & 27.2 $\pm$ 0.9 & $>$14.10 \\ 
{\SiIV} & -143 $\pm$ 1 & 7.1 $\pm$ 1.4 & 12.56 $\pm$ 0.05 \\ 
{\SiIV} & -111 $\pm$ 2 & 11.3 $\pm$ 2.9 & 12.43 $\pm$ 0.08 \\ 
{\SiIV} & -24 $\pm$ 1 & 20.0 $\pm$ 0.4 & $>$13.74 \\ 
{\SiII} & -145 $\pm$ 1 & 2.4 $\pm$ 0.5 & 12.98 $\pm$ 0.09 \\ 
{\SiII} & -130 $\pm$ 1 & 16.9 $\pm$ 0.7 & 13.08 $\pm$ 0.02 \\ 
{\SiII} & -61 $\pm$ 1 & 10.5 $\pm$ 1.6 & 12.72 $\pm$ 0.06 \\ 
{\SiII} & -7 $\pm$ 1 & 20.4 $\pm$ 0.5 & $>$14.98  \\ 
{\CII} & -17 $\pm$ 1 & 23.1 $\pm$ 0.6 & $>$15.87  \\ 
{\CII} & -129 $\pm$ 1 & 17.7 $\pm$ 0.5 & $>$14.15  \\ 
{\CII} & -143 $\pm$ 1 & 4.3 $\pm$ 1.1 & $>$13.86  \\ 
{\CIV} & -146 $\pm$ 1 & 6.0 $\pm$ 1.7 & 13.14 $\pm$ 0.09 \\ 
{\CIV} & -113 $\pm$ 2 & 22.7 $\pm$ 2.5 & 13.69 $\pm$ 0.04 \\ 
{\CIV} & -26 $\pm$ 1 & 29.4 $\pm$ 0.8 & $>$14.33  \\ 
{\AlII} & -143 $\pm$ 2 & 5.4 $\pm$ 2.8 & 11.79 $\pm$ 0.12 \\ 
{\AlII} & -129 $\pm$ 3 & 0.3 $\pm$ 4.2 & 12.12 $\pm$ 0.81 \\ 
{\AlII} & -56 $\pm$ 2 & 7.2 $\pm$ 3.5 & 11.69 $\pm$ 0.13 \\ 
{\AlII} & -7 $\pm$ 1 & 16.5 $\pm$ 1.4 & $>$13.96  \\ 
{\FeII} & -143 $\pm$ 2 & 5.4 $\pm$ 2.8 & 11.79 $\pm$ 0.12 \\ 
{\FeII} & -129 $\pm$ 3 & 0.3 $\pm$ 4.2 & 12.12 $\pm$ 0.81 \\ 
{\FeII} & -56 $\pm$ 2 & 7.2 $\pm$ 3.5 & 11.69 $\pm$ 0.13 \\ 
{\FeII} & -7 $\pm$ 1 & 16.5 $\pm$ 1.4 & $>$13.96  \\ 
\hline
        &                       &       MRK1392         &             $\langle R \rangle$= 6.5 kpc              \\
\hline
{\SiIII} & -86 $\pm$ 24 & 54.1 $\pm$ 21.4 & 12.71 $\pm$ 0.23 \\ 
{\SiIII} & -14 $\pm$ 2 & 35.3 $\pm$ 1.7 & $>$13.57  \\ 
{\SiIV} & -100 $\pm$ 6 & 21.2 $\pm$ 8.8 & 12.40 $\pm$ 0.13 \\ 
{\SiIV} & -21 $\pm$ 1 & 28.5 $\pm$ 0.9 & $>$13.71 \\ 
{\SiII} & -120 $\pm$ 2 & 15.8 $\pm$ 0.6 & 12.64 $\pm$ 0.47 \\ 
{\SiII} & -88 $\pm$ 1 & 14.2 $\pm$ 1.8 & 12.69 $\pm$ 0.03 \\ 
{\SiII} & -12 $\pm$ 1 & 25.6 $\pm$ 0.5 & $>$14.38  \\ 
{\CII} & -9 $\pm$ 1 & 37.3 $\pm$ 1.1 & $>$14.80 \\ 
{\CII} & -83 $\pm$ 1 & 7.5 $\pm$ 2.2 & 13.68 $\pm$ 0.05 \\ 
{\CII} & -117 $\pm$ 2 & 19.4 $\pm$ 2.9 & 13.74 $\pm$ 0.05 \\ 
{\CIV} & -67 $\pm$ 18 & 127.8 $\pm$ 22.5 & 13.54 $\pm$ 0.08 \\ 
{\CIV} & -28 $\pm$ 1 & 27.7 $\pm$ 1.0 & $>$14.20  \\ 
{\AlII} & -93 $\pm$ 3 & 6.3 $\pm$ 7.5 & 11.80 $\pm$ 0.14 \\ 
{\AlII} & -16 $\pm$ 1 & 29.8 $\pm$ 1.4 & 13.37 $\pm$ 0.03 \\ 
{\FeII} & -93 $\pm$ 3 & 6.3 $\pm$ 7.5 & 11.80 $\pm$ 0.14 \\ 
{\FeII} & -16 $\pm$ 1 & 29.8 $\pm$ 1.4 & $>$13.37  \\ 
\hline
        &                       &       SDSSJ151237.15+012846.0         &             $\langle R \rangle$= 6.31 kpc              \\
\hline
{\SiIII}   &  -114 $\pm$ 5 & 30.3 $\pm$ 6.7 & 13.3 $\pm$ 0.1  \\
{\SiIII}   &  -15 $\pm$ 3 & 19.5 $\pm$ 16.7 &  $>$15.0  \\
{\CII} & -114 $\pm$ 9 & 25.0 $\pm$ 10.0 & 13.65 $\pm$ 0.30 \\ 
{\CII} & 4 $\pm$ 7 & 50.0 $\pm$ 12.0 & $>$14.80  \\ 
{\SiII} & -119 $\pm$ 3 & 7.0 $\pm$ 10.2 & 12.93 $\pm$ 0.18 \\ 
{\SiII} & -3 $\pm$ 4 & 37.9 $\pm$ 6.1 & $>$14.22  \\ 
{\SiIV}   &  -112 $\pm$ 9 & 29.0 $\pm$ 12.8 & 13.1 $\pm$ 0.2 \\  
{\SiIV}   &  1 $\pm$ 3 & 37.5 $\pm$ 4.7 & $>$13.8  \\ 
\enddata
\tablenotetext{a}{The random velocity error from the profile fit process is listed.  The actual velocity error must also include the $\pm$5 {\kms} COS velocity calibration error.}
\tablenotetext{b}{Saturated lines are indicated with $>$ and give the lower limits on $\log N$.}
\label{table:Vpfit_measurements}
\end{deluxetable*}

\clearpage{}
\begin{deluxetable*}{cccccc}
\tablecolumns{5}
\tablewidth{0pt}
\tablecaption{Minimum mass outflow rate estimates}
\tablehead{
\colhead{ QSO Name }&
\colhead{ $\dot{M}_{out}$ [$\rm{ M_{\odot}\; yr^{-1}}$]\tablenotemark{a}} &
\colhead{ $v_{r}$ [km$\rm{s^{-1}}$]\tablenotemark{b}}&
\colhead{ $R \;$ [kpc]\tablenotemark{c}}&
\colhead{$\log N_{H} \rm{[cm^{-2}]}\; $} &
\colhead{$\dot{P}_{out} \text{[dynes]}$\tablenotemark{d}}
}
\startdata
PDS 456            &     $\gtrsim$ 0.20  &     479  &     2.27  & $\gtrsim$ 19.1   &$\gtrsim$ 6 $\times 10^{32}$ \\
1H1613-097       &     $\gtrsim$ 0.28  &      410  &      4.06  &  $\gtrsim$ 19.04 & $\gtrsim$ 7.2$\times 10^{32}$ \\
\enddata
\tablenotetext{b}{ Minimum mass outflow rates for the two lowest latitude lines of sight}
\tablenotetext{b}{ Radial outflow velocity from model M1 at a radial distance $R$.}
\tablenotetext{c}{Mean radial distance ($R$) from the GC.}
\tablenotetext{d}{Minimum momentum flux carried out by the outflowing gas from the GC.}
\label{mass_outflow_rate}
\end{deluxetable*}

\clearpage{}
\appendix
In the appendix we present the HST-COS spectra of the 42 QSO sightlines outside the northern Fermi Bubble within 35 degrees longitude of the GC.


\renewcommand\bibsection{}
\bibliographystyle{apj}
\bibliography{mybibliography}

\clearpage

\begin{deluxetable*}{lllll lllll l}
\tabletypesize{\small}
\tablewidth{0pt}
\tablecaption{HVC Detections in the Northern Fermi Bubble Region\tablenotemark{a}}
\tablehead{Sightline & $l$ & $b$ & $v_-$\tablenotemark{b} & $v_+$\tablenotemark{b} & 
$\langle v_{\rm LSR} \rangle$\tablenotemark{c} & $ \langle v_{\rm dev} \rangle$\tablenotemark{d} & Line &
log\,$N_{\rm a}$\tablenotemark{e} & Location & Notes\\
 & (\degr) & (\degr) & & &  &  & & ($N_{\rm a}$ in cm$^{-2}$) & & } 
\startdata 
SDSSJ141038.40+230447.0 & 24.6 & 71.6 & 135 & 200 & 166 & 163 & \ion{C}{2}  $\lambda$1334 & 13.56$^{+0.09}_{-0.12}$ & Outside & Complex K?\\
                        &      &      &     &     &     &     & \ion{Si}{3} $\lambda$1206 & 12.39$^{+0.12}_{-0.16}$ &         &          \\

        RXJ1426.2+1955  & 19.6 & 67.2 & $-$150 &  $-$95 & $-$113 & $-$113 & \ion{C}{2}  $\lambda$1334 & 13.97$^{+0.05}_{-0.05}$ & Outside & Complex K?\\
                        &      &      &        &        &        &        & \ion{Si}{2} $\lambda$1260 & 12.48$^{+0.08}_{-0.10}$ &         &    \\
                        &      &      &        &        &        &        & \ion{Si}{3} $\lambda$1206 & 12.36$^{+0.10}_{-0.14}$ &         &    \\

SDSSJ135726.27+043541.4 & 340.8 & 62.5 & $-$160 &  $-$75 & $-$115 & $-$110 & \ion{C}{2}  $\lambda$1334 & 14.06$^{+0.04}_{-0.04}$ & Outside & ...\\
                        &      &      &        &        &        &        & \ion{Si}{2} $\lambda$1260 & 12.92$^{+0.04}_{-0.05}$ &         &    \\
                        &      &      &        &        &        &        & \ion{Si}{3} $\lambda$1206 & $>$12.90  &         &    \\
                        &      &      &        &        &        &        & \ion{C}{4}  $\lambda$1548 & 13.60$^{+0.06}_{-0.06}$ &         &    \\
                        &      &      &        &        &        &        & \ion{C}{4}  $\lambda$1550 & 13.48$^{+0.12}_{-0.16}$ &         &    \\

SDSSJ141949.40+060654.0 & 351.9 & 60.3 & $-$160 &  $-$75 & $-$107 & $-$104 & \ion{C}{2}  $\lambda$1334 & $>$14.25              & Outside & ...\\
                        &      &      &        &        &        &        & \ion{Si}{2} $\lambda$1260 & 12.97$^{+0.07}_{-0.08}$ &         &    \\
                        &      &      &        &        &        &        & \ion{Si}{2} $\lambda$1193 & 13.05$^{+0.12}_{-0.17}$ &         &    \\
                        &      &      &        &        &        &        & \ion{Si}{2} $\lambda$1190 & 13.27$^{+0.14}_{-0.20}$ &         &    \\
                        &      &      &        &        &        &        & \ion{Si}{3} $\lambda$1206 & 12.95$^{+0.07}_{-0.08}$ &         &    \\
                        &      &      &        &        &        &        & \ion{C}{4}  $\lambda$1548 & $>$13.52              &         &    \\
                        &      &      &        &        &        &        & \ion{C}{4}  $\lambda$1550 & $<$13.86              &         & \\

            HE1340-0038 & 328.8 & 59.4 & $-$160 & $-$70 & $-$107 & $-$99   & \ion{C}{2}  $\lambda$1334 & 14.22$^{+0.03}_{-0.03}$ & Outside & ...\\
                        &      &      &        &       &        &         & \ion{O}{1}  $\lambda$1302 & 14.13$^{+0.08}_{-0.10}$ &         &    \\
                        &      &      &        &       &        &         & \ion{Si}{2} $\lambda$1260 & 13.15$^{+0.03}_{-0.05}$ &         &    \\
                        &      &      &        &       &        &         & \ion{Si}{2} $\lambda$1193 & 13.53$^{+0.04}_{-0.05}$ &         &    \\ 
                        &      &      &        &       &        &         & \ion{Si}{2} $\lambda$1190 & 13.31$^{+0.09}_{-0.11}$ &         &    \\
                        &      &      &        &       &        &         & \ion{Si}{2} $\lambda$1526 & 13.69$^{+0.10}_{-0.12}$ &         &    \\
                        &      &      &        &       &        &         & \ion{Si}{3} $\lambda$1206 & 12.92$^{+0.04}_{-0.05}$ &         &    \\
 
                 MRK841 & 11.2 & 54.6 &    155 &   235 &    198 &    193  & \ion{C}{2}  $\lambda$1334 & 13.91$^{+0.04}_{-0.04}$ & Outside & East of Complex K\\
                        &      &      &        &       &        &         & \ion{C}{2}$^*$  $\lambda$1335 & 13.26$^{+0.13}_{-0.20}$ &         &    \\
                        &      &      &        &       &        &         & \ion{O}{1}  $\lambda$1302 & 13.78$^{+0.10}_{-0.13}$ &         &    \\
                        &      &      &        &       &        &         & \ion{Si}{2} $\lambda$1260 & 12.62$^{+0.04}_{-0.04}$ &         &    \\
                        &      &      &        &       &        &         & \ion{Si}{2} $\lambda$1193 & 12.79$^{+0.10}_{-0.13}$ &         &    \\ 
                        &      &      &        &       &        &         & \ion{Si}{2} $\lambda$1190 & 13.03$^{+0.11}_{-0.15}$ &         &    \\
                        &      &      &        &       &        &         & \ion{Si}{2} $\lambda$1526 & 13.14$^{+0.14}_{-0.21}$ &         &    \\
                        &      &      &        &       &        &         & \ion{Si}{3} $\lambda$1206 & 12.37$^{+0.09}_{-0.12}$ &         &    \\

SDSSJ150928.30+070235.0 &  7.8 & 51.6 & $-$150 & $-$90 & $-$114 & $-$114  & \ion{C}{2}  $\lambda$1334 & $>$14.29 & Boundary & East of Complex K \\
                        &      &      &        &       &        &         & \ion{O}{1}  $\lambda$1302 & $>$14.40 &         &    \\
                        &      &      &        &       &        &         & \ion{Si}{2} $\lambda$1260 & 12.98$^{+0.06}_{-0.07}$ &         &    \\
                        &      &      &        &       &        &         & \ion{Si}{2} $\lambda$1193 & 13.12$^{+0.09}_{-0.12}$ &         &    \\ 
                        &      &      &        &       &        &         & \ion{Si}{2} $\lambda$1190 & 13.26$^{+0.12}_{-0.17}$ &         &    \\
                        &      &      &        &       &        &         & \ion{Si}{3} $\lambda$1206 & $>$13.11 &         &    \\
                        &      &      &        &       &        &         & \ion{Si}{4} $\lambda$1393 & 12.74$^{+0.16}_{-0.25}$ &         &    \\

                RBS1454 &  5.6 & 52.9 & $-$140 & $-$75 & $-$103 & $-$103 & \ion{C}{2}  $\lambda$1334 & 13.73$^{+0.08}_{-0.09}$ & Boundary & East of Complex K \\
                        &      &      &        &        &        &        & \ion{Si}{2} $\lambda$1260 & 12.66$^{+0.10}_{-0.14}$ &         &    \\
                        &      &      &        &        &        &        & \ion{Si}{3} $\lambda$1206 & 12.75$^{+0.07}_{-0.08}$ &         &    \\

SDSSJ150952.20+111047.0 & 13.6 & 53.8 &   195  &  265   &   232 &  225    & \ion{C}{2}$^*$  $\lambda$1335 & 13.65$^{+0.08}_{-0.09}$ & Outside & East of Complex K\\
                        &      &      &        &       &        &         & \ion{O}{1}  $\lambda$1302 & 14.39$^{+0.08}_{-0.09}$ &         &    \\
                        &      &      &        &       &        &         & \ion{Si}{2} $\lambda$1260 & $>$13.19 &         &    \\
                        &      &      &        &       &        &         & \ion{Si}{2} $\lambda$1193 & $>$13.33 &         &    \\ 
                        &      &      &        &       &        &         & \ion{Si}{2} $\lambda$1190 & $<$13.24 &         &    \\
                        &      &      &        &       &        &         & \ion{Si}{3} $\lambda$1206 & 12.83$^{+0.08}_{-0.10}$ &         &    \\

  SDSSJ154553.50+093620.0 & 18.3 & 45.4 & $-$175 & $-$125 & $-$145 & $-$145 & \ion{Al}{2} $\lambda$1670 & $>$12.54 & Boundary & East of Complex K \\
                          &        &      &        &        &        &        & \ion{Si}{2} $\lambda$1526 & 13.67$^{+0.09}_{-0.12}$ &         &    \\
  
             PG1522+101 & 14.9 & 50.1 & $-$135 & $-$95 & $-$108 & $-$108  & \ion{C}{2}  $\lambda$1334 & 13.70$^{+0.05}_{-0.06}$ & Boundary & East of Complex K\\
                        &      &      &        &       &        &         & \ion{O}{1}  $\lambda$1302 & 13.89$^{+0.13}_{-0.18}$ &         &    \\
                        &      &      &        &       &        &         & \ion{Si}{2} $\lambda$1260 & 12.68$^{+0.05}_{-0.06}$ &         &    \\
                        &      &      &        &       &        &         & \ion{Si}{2} $\lambda$1193 & 12.75$^{+0.09}_{-0.11}$ &         &    \\ 
                        &      &      &        &       &        &         & \ion{Si}{2} $\lambda$1190 & 13.12$^{+0.08}_{-0.10}$ &         &    \\
                        &      &      &        &       &        &         & \ion{Si}{2} $\lambda$1526 & 13.17$^{+0.10}_{-0.13}$ &         &    \\
                        &      &      &        &       &        &         & \ion{Si}{3} $\lambda$1206 & 12.54$^{+0.06}_{-0.07}$ &         &    \\
                        &      &      &        &        &        &        & \ion{C}{4}  $\lambda$1548 & 12.96$^{+0.12}_{-0.16}$ &         &    \\
                        &      &      &        &        &        &        & \ion{C}{4}  $\lambda$1550 & $<$13.02              &         &    \\
       QSO1500-4140     & 327.7 & 14.6 & 120 & 220 &   +174 & $-$174 & \ion{C}{2}  $\lambda$1334 & $>$14.33 & Outside & ...\\
                        &      &      &        &        &        &        & \ion{Si}{2} $\lambda$1260 & $>$13.11 &         &    \\
                        &      &      &        &        &        &        & \ion{Si}{2} $\lambda$1260 & $>$13.11 &         &    \\
                        &      &      &        &        &        &        & \ion{Si}{3} $\lambda$1206 & $>$13.07 &         &    \\

\enddata
\vspace{-0.1cm}
\label{appendix_table}
\end{deluxetable*}
\vspace{-0.1cm}

$^a$ Northern FB region is defined as $320\degr\!<\!l\!<\!40\degr$, $b\!>\!0\degr$ to include the FB and its surroundings. Sightlines are presented
in order of decreasing latitude. STIS sightline M5-ZNG1 is not included here; see Table 2.
$^b$ Minimum and maximum LSR velocity of absorption component.
$^c$ Central LSR velocity of component.
$^d$ Deviation velocity of component, defined as the offset from a model of cylindrical co-rotation. 
$^e$ Apparent column density \citep{Savage1991} integrated between $v_-$ and $v_+$. Errors include statistical and continuum-placement uncertainties. Lower limits are given for saturated absorbers. 
We are excluding the MRK 1383 sightline in this study as that sightline does not have a HST/COS spectrum.

\begin{figure*}\setcounter{figure}{0} \renewcommand{\thefigure}{A.\arabic{figure}} 
\centering
\subfloat{\label{fig:appendix}\includegraphics[height=20. cm,width=1.\textwidth]{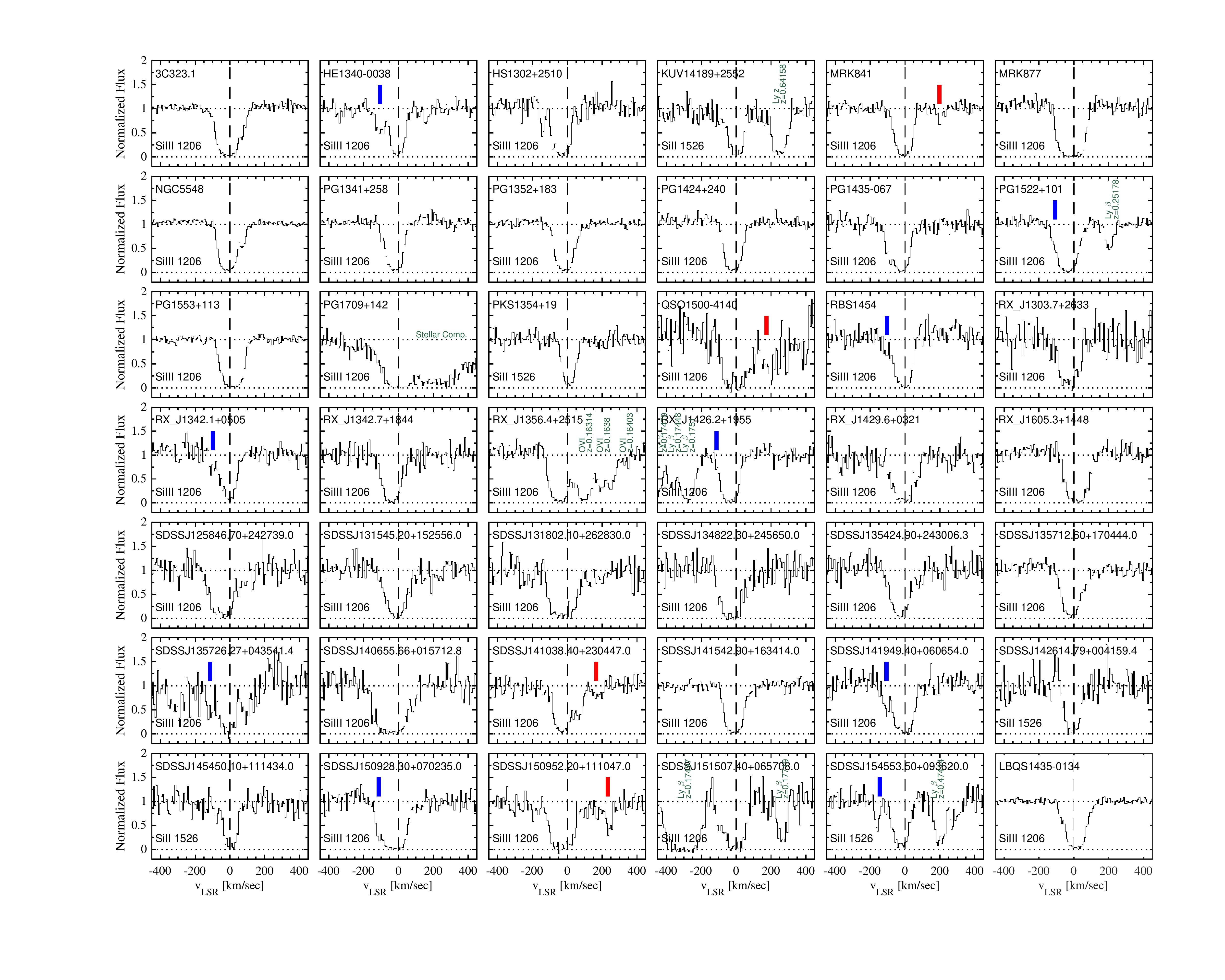}}
\qquad
\caption{HST-COS  spectra of the {\SiIII} and {\CII} transitions for lines of sight outside the northern Fermi Bubble. The vertical  ticks indicate the centroids of individual Voigt profile components. Both the  redshifted components (red ticks) and blueshifted components (blue ticks) are flagged. For sightlines with COS/G160M data only, we show the corresponding  {\SiII} and {\CIV} transitions, respectively.}
\end{figure*}
\clearpage
\begin{figure*}\setcounter{figure}{1} \renewcommand{\thefigure}{A.\arabic{figure}} 
\ContinuedFloat
\flushleft
\subfloat{\label{fig:appendix}\includegraphics[height=20 cm,width=1.\textwidth]{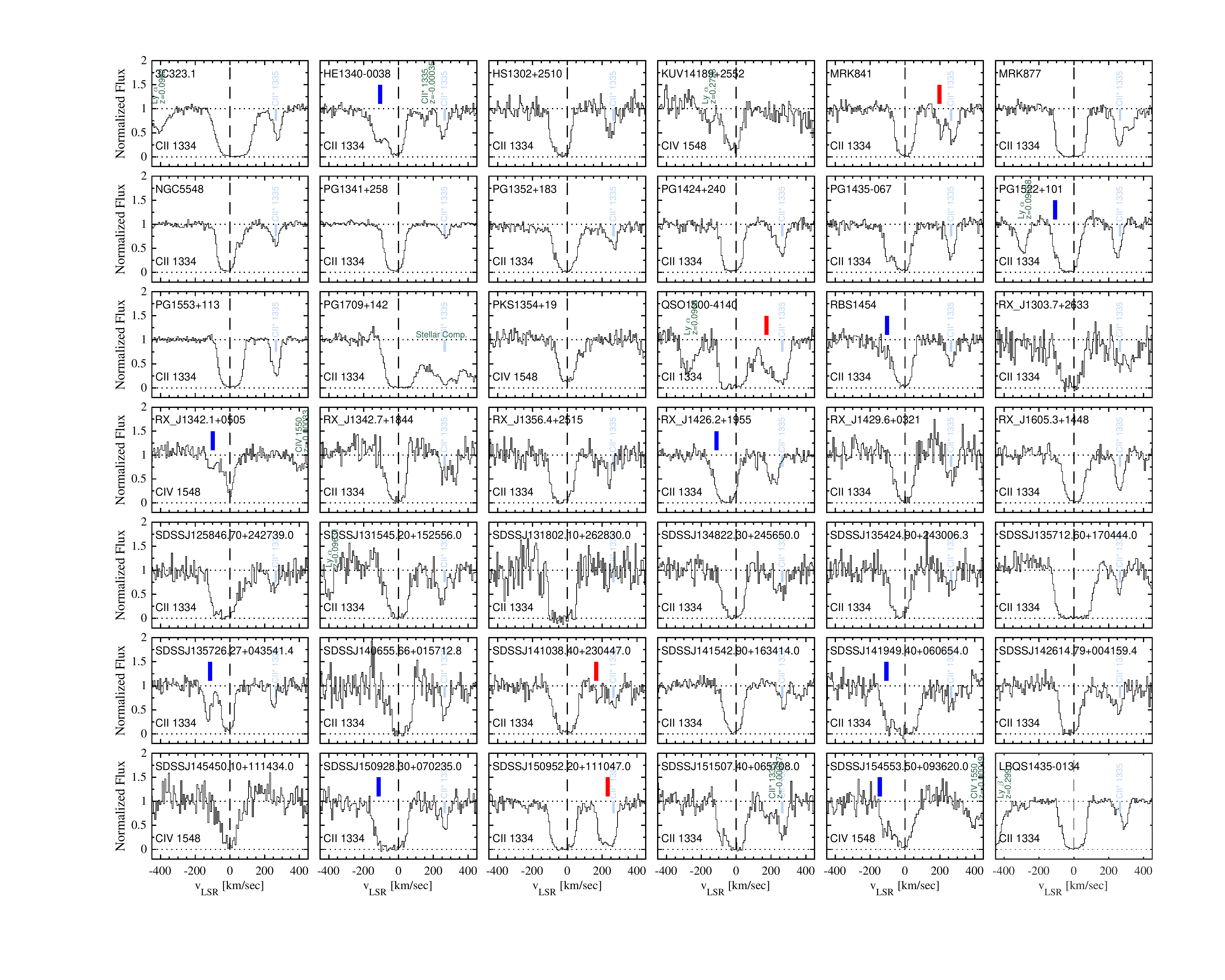}}
\qquad
\caption{continued}
\end{figure*}

\begin{figure*}\setcounter{figure}{1} \renewcommand{\thefigure}{A.\arabic{figure}} 
\centering
\includegraphics[width=0.8\textwidth]{{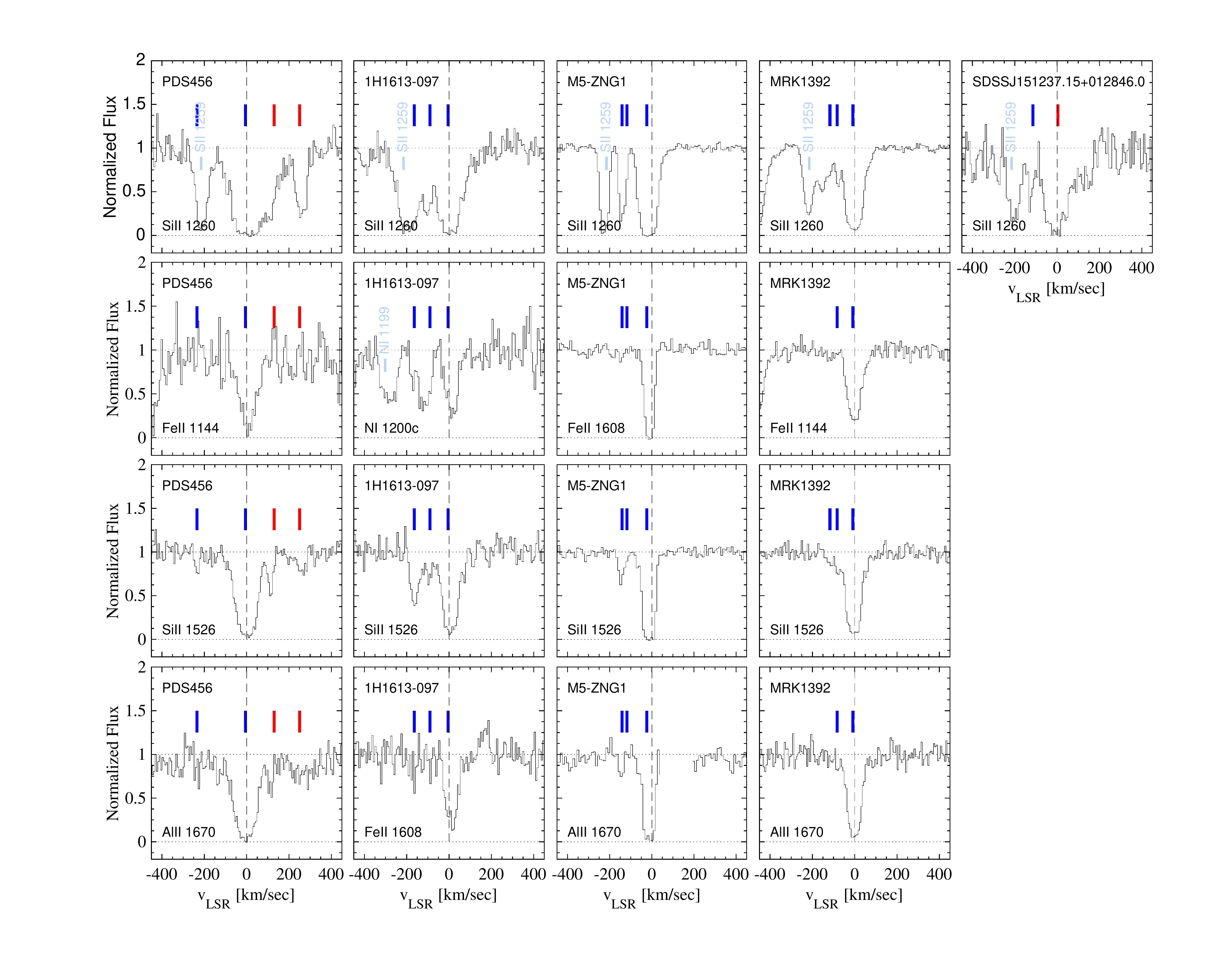}}
\caption{HST-COS  spectra of additional transitions for the five lines of sight inside the northern Fermi Bubble. The vertical  ticks indicate the velocities of HVCs along these directions. Both the  redshifted components (red ticks) and blueshifted components (blue ticks) are flagged.}
\label{fig:appendix2}
\end{figure*}

\end{document}